\documentclass[letterpaper]{svjour3}
\smartqed
\usepackage{graphicx}
\usepackage{mathptmx}
\usepackage{url}
\usepackage{aas_macros}
\usepackage{natbib}
\usepackage[letterpaper,breaklinks]{hyperref}
\bibpunct{(}{)}{;}{a}{}{,}

\def\makeheadbox{{%
\hbox to0pt{\hbox to\hsize{\vbox{%
\hbox{Space Sci Rev}
\hbox{\href{http://dx.doi.org/10.1007/s11214-008-9473-6}{DOI 10.1007/s11214-008-9473-6}}
\hbox{The original publication is available at www.springerlink.com.}
}\hfil}%
\hss}}}

\newcommand\arcsec{\mbox{$^{\prime\prime}$}}
\newcommand\CaIIH{\mbox{Ca\,\textsc{ii}\,H}}

\newcommand\CaI{\mbox{Ca\,\textsc{i}}}
\newcommand\FeI{\mbox{Fe\,\textsc{i}}}
\newcommand\SrI{\mbox{Sr\,\textsc{i}}}
\newcommand\MgI{\mbox{Mg\,\textsc{i}}}
\newcommand\MnI{\mbox{Mn\,\textsc{i}}}
\newcommand\Halpha{\mbox{H\,$\alpha$}}
\newcommand\CC{\mbox{C$_2$}}

\newcommand\nm{\ensuremath{\mathrm{nm}}}
\newcommand\mum{\ensuremath{\mu\mathrm{m}}}
\newcommand\sqcm{\ensuremath{\mathrm{cm}^2}}
\newcommand\m{\ensuremath{\mathrm{m}}}
\newcommand\km{\ensuremath{\mathrm{km}}}
\newcommand\kms{\ensuremath{\mathrm{km}/\mathrm{s}}}
\newcommand\cms{\ensuremath{\mathrm{cm}/\mathrm{s}}}

\newcommand\gccm{\ensuremath{\mathrm{g}/\mathrm{cm}^3}}
\newcommand\gauss{\ensuremath{\mathrm{G}}}
\newcommand\kgauss{\ensuremath{\mathrm{kG}}}
\newcommand\mxsqcm{\ensuremath{\mathrm{Mx}/\mathrm{cm}^2}}
\newcommand\mx{\ensuremath{\mathrm{Mx}}}
\newcommand\kelvin{\ensuremath{\mathrm{K}}}

\begin{document}

\title{Small-scale solar magnetic fields}

\author{A.~G.~de Wijn\and
J.~O.~Stenflo\and
S.~K.~Solanki\and
S.~Tsuneta}

\institute{A.~G.~de Wijn\at
University Corporation for Atmospheric Research, P.O.~Box 3000, Boulder, CO 80307-3000, USA\\
\email{dwijn@ucar.edu}%
\and
J.~O.~Stenflo\at
Institute of Astronomy, ETH Zurich, CH-8093 Zurich, Switzerland\\
\email{stenflo@astro.phys.ethz.ch}%
\and
S.~K.~Solanki\at
Max-Planck-Institut f\"ur Sonnensystemforschung, Max-Planck-Strasse~2, 37191 Katlenburg-Lindau, Germany\\
\email{solanki@mps.mpg.de}%
\and
S.~Tsuneta\at
National Astronomical Observatory, Mitaka, Tokyo 181-8588, Japan\\
\email{saku.tsuneta@nao.ac.jp}%
}

\journalname{Space Sci Rev}

\date{Received: 12 August 2008 / Accepted: 26 November 2008}

\maketitle

\begin{abstract}
As we resolve ever smaller structures in the solar atmosphere, it has become clear that magnetism is an important component of those small structures.
Small-scale magnetism holds the key to many poorly understood facets of solar magnetism on all scales, such as the existence of a local dynamo, chromospheric heating, and flux emergence, to name a few.
Here, we review our knowledge of small-scale photospheric fields, with particular emphasis on quiet-sun field, and discuss the implications of several results obtained recently using new instruments, as well as future prospects in this field of research.
\keywords{sun \and photosphere \and magnetism \and small scale}
\end{abstract}

\section{Introduction}
\label{sec:intro}
Magnetism on the sun occurs on all scales.
It manifests itself at the largest scales as a mean-field component that covers an entire hemisphere, and on progressively smaller scales as active regions, sunspots, and pores.
Magnetic field in the lower solar atmosphere has structure on the smallest observable scales, up to the diffraction limit of the best telescopes.
Theoretical arguments and simulations indicate that there is structure well beyond what can be observed today or in the forseeable future.

A comprehensive ab-initio model of magnetic activity is currently impossible from a practical standpoint, and will remain so in the near future.
The complex interaction of magnetic field, hydrodynamics, and radiative transfer requires sophisticated numerical analysis.
A simulation would have to cover a substantial surface area over a good fraction of the convection zone in order to capture large-scale patterns such as supergranulation, yet also have sufficient resolution to capture interactions on scales of several kilometers or less.
Such a simulation is prohibitively expensive in terms of computation time.
In order to gain understanding of the physical processes involved in the creation, evolution, and eventual destruction of magnetic field, we must turn to observations to study the properties of magnetic structures.

Granular flows in the photosphere expunge field from cell interiors.
Flux is swept into the intergranular lanes, where it clumps in small concentrations of mostly vertical field with strengths in excess of one kilogauss.
Bright points and faculae, the most conspicuous features of magnetism in the lower solar atmosphere, correspond to these small concentrations of field.
They are well-known in active regions, where they group together in plages.
In the quiet sun, supergranular flows concentrate them in the magnetic network that incompletely outlines supergranular cells.

Internetwork quiet-sun magnetism has been somewhat ignored historically, largely due to a lack of observations with sufficient resolution and accuracy.
However, it has attracted particular interest in the past years, and this subject is currently being studied vigorously.
Isolated concentrations of strong field that produce bright points and faculae also exist in supergranular interiors.
Outside the concentrations, much weaker field that is not predominantly oriented perpendicular to the surface is ubiquitously present.
This more horizontal field typically does not produce bright points that are easily observed using proxy-magnetometry diagnostics such as imaging in the Fraunhofer G band.
Instead, sensitive magnetometers are required to observe and study weak field.
The development of new instrumentation and seeing-mitigating techniques (the SpectroPolarimeter instrument on the space-borne observatory Hinode is an excellent example of both), and advanced simulations facilitated by the steadily increasing processing power of computers have made it possible to study this subject in detail.

Here, we focus our attention on magnetic fine structure of the quiet solar photosphere.
In particular, we will discuss internetwork field.
Quiet-sun internetwork areas cover the majority of the solar surface.
Four orders of magnitude more flux emerges in the internetwork than in active regions.
Consequently, field in these areas is of importance in understanding certain aspects of solar magnetic activity, such as the existence and workings of a granular dynamo and the dynamical coupling of the photosphere to higher layers.

We aim to provide a comprehensive review of quiet-sun internetwork magnetic fine structure, starting with a general overview in Sect.~\ref{sec:qsmf}.
We then address several small-scale phenomena that have recently attracted particular attention as a result of new, high-resolution observations: properties of horizontal field in the photosphere (Sect.~\ref{sec:thmf}), polar field (Sect.~\ref{sec:pf}), and concentrations of strong vertical field (Sect.~\ref{sec:bpme}).
Section~\ref{sec:urmf} concludes the chapter with a discussion on unresolved fields.

\section{Quiet Sun magnetic fields}
\label{sec:qsmf}
The magnetic field found in the quiet sun can be categorized into network and internetwork field.
The latter were discovered by
	\cite{1971IAUS...43...51L,Livingston1975}
on the basis of a weak Stokes $V$ signal found in the interiors of supergranular cells.
A separate category, the turbulent field, has also been proposed.
It is not clear which of the further proposed types of quiet-sun magnetic fields, such as `granular fields'
	\citep{1999ApJ...514..448L},
horizontal quiet-sun fields
	\citep{1996ApJ...460.1019L},
or `seething fields'
	\citep{2007ApJ...659L.177H}
describe independent types of magnetic structures, and which are just different names for the same physical entity, detected in different types of observational data.
The different techniques used to detect and study them pose the main difficulty with identifying one with the other.
E.g., internetwork fields have traditionally been measured using the Zeeman effect, while the turbulent field has been probed mainly through the Hanle effect.
Because of the cancellation of the Zeeman signal in the presence of opposite-polarity longitudinal fields in the resolution element, a tangled field may largely escape detection through the Zeeman effect, especially if the field is intrinsically weak.
Only the larger scales of such a turbulent field would be seen using Zeeman-based diagnostics.
These may then appear like internetwork elements, which in this scenario would represent just the tip of the iceberg of the sun's turbulent field.

There have also been suggestions that the magnetic fluxes of all magnetic features in the photosphere form similar patterns irrespective of the scale at which they are observed.
This scale invariance is consistent with the proposal that the magnetic field forms a fractal (or multifractal) pattern at the solar surface
	\cite[e.g.,][]{1987SoPh..107...11R,
	1995PhRvE..51..316L,
	1995SoPh..157...45K,
	1996A&A...308..213N,
	1999ApJ...515..801M,
	2004A&A...420..333M,
	2002ESASP.505..101S,
	2003ASPC..286..169S,
	2005SoPh..228...29A,
	2007A&A...461..331C}.
Since magnetic features are moved around by the evolving convection cells, possibly such an analysis provides more information on the distribution of convection at different scales, rather than on intrinsic magnetic properties.
Convective eddies are expected to be self-similar for a turbulent medium, such as the solar convection zone.

\subsection{Magnetic flux in the quiet sun}

\subsubsection{Methods}

In principle, it is possible to detect magnetic features and partly to estimate their magnetic flux in a variety of ways.
However, the different types of measurements give different results, so that some uncertainty remains on just how much magnetic flux the quiet sun harbors.

Contrasts in more or less narrow wavelength bands are widely used as proxies of the magnetic field, since they are easy to observe at high resolution even under variable seeing conditions.
They include the brightness in the G band, \CaIIH\ or K line core, or CN bandhead.
These proxies are, however, not ideal for determining the magnetic flux in the quiet sun, due to their small sensitivity.
There is still some uncertainty to what extent internetwork magnetic features produce visible signatures in these proxies
	\citep[however, see][]{2005A&A...441.1183D}.

The Zeeman effect not only provides quantitative measurements of the magnetic vector, but is also much more sensitive to small amounts of magnetic flux and has been shown to sense fluxes as low as $10^{16}~\mx$ (or even less), particularly if the field is aligned along the line of sight (i.e., well visible in Stokes $V$, the net circular polarization).
It suffers, however, from the fact that Stokes $V$ is also sensitive to the direction in which the flux points (towards or away from the observer), so that if there is a mixture of polarities on a sufficiently small scale, the signal in Stokes $V$ can be canceled.
In Stokes $Q$ and $U$ cancellation, although possible, is less likely (it requires two transverse fields at right angles to each other in the resolution element).

If the aim is to measure intrinsically weak, possibly turbulent fields, then the Hanle effect is the method of choice.
Basically, the Hanle effect allows the magnetic vector to be determined if the field strength lies within a fiducial range that depends on the observed spectral line.
The Hanle effect is generally sensitive to low intrinsic field strengths (typical values are below a few $100~\gauss$, depending on the spectral line).
Of importance for the field in the quiet sun is that the Hanle effect allows a weighted average of the field strength to be obtained even for a field that is isotropically distributed in the resolution element.
Such a field would be invisible to the Zeeman effect as long as it doesn't produce any significant broadening of the line profiles (see below).

\subsubsection{Measurements of magnetic flux in the quiet sun}

The flux distribution in network elements has been determined by, e.g.,
	\cite{1998A&A...331..771M}
and
	\cite{2001ApJ...555..448H}
	\citep[cf.][]{1997ApJ...487..424S}.
They all find an exponential increase in the number density of elements with decreasing flux, down to the sensitivity limit (lying at $2\times10^{18}~\mx$ for the investigation of
	\citealp{2001ApJ...555..448H}).
In contrast to this result,
	\cite{1995SoPh..160..277W}
obtain a non-exponential, non-power law distribution for the network fluxes and a different (but also non-exponential, non-power law) distribution for the internetwork field.
They use a series of criteria to differentiate between the two, including location (at the edges of supergranules or in their interior), proper motion speeds (higher speed of internetwork elements), etc.
The weakest fluxes of individual internetwork features that they record are $10^{16}~\mx$.

	\cite{1987SoPh..110..101Z}
found that the rate of magnetic flux emergence in internetwork fields is roughly 100 times larger than in ephemeral active regions.
In the latter it is another 100 times higher than in normal active regions.
Therefore, the internetwork fields completely dominate the flux emergence.
However, whether the internetwork fields dominate the total flux at any given time depends on the ratio of emergence time scale to decay time scale of the fields.
Intranetwork fields not just emerge at the highest rate, but also decay the most rapidly, so that their exact contribution to the instantaneous total
magnetic flux is still unclear.

Prior to the Hinode mission
	\citep{2007SoPh..243....3K},
the typical average field strength in the quiet sun obtained from Zeeman effect measurements were a few gauss (typically $2$--$5~\gauss$).
The estimates of
	\cite{2003ApJ...582L..55D,
	2003A&A...407..741D}
and
	\cite{2005A&A...436L..27K}
	\cite[cf.][]{2005A&A...442.1059K}
count as exceptions.
At a spatial resolution of $0.5\arcsec$,
	\cite{2003ApJ...582L..55D,
	2003A&A...407..741D}
obtained an average field strength of $20~\gauss$ in the internetwork.
	\cite{2005A&A...436L..27K}
compared the distributions of Stokes $V$ amplitudes simultaneously observed in the infrared and the visible with the amplitudes of synthetic profiles computed in snapshots of mixed-polarity 3D MHD simulations harboring different amounts of magnetic flux.
The MHD simulations of
	\cite{2005A&A...429..335V}
can be studied at a spatial resolution nearly an order or magnitude higher than the
observations, so that mixed polarity magnetic flux that is canceled out in
the observations can still be counted in the simulations.
The magnetic flux in the simulation snapshot that gives the best fit to the Stokes $V$ amplitude distributions of both, the infrared and the visible lines, is taken to represent the solar flux.
In this manner, to first order, the problem that a part of the flux is canceled within each spatial resolution element of the observations is circumvented.
Just like
	\cite{2003ApJ...582L..55D,
	2003A&A...407..741D},
	\cite{2005A&A...436L..27K}
also obtained $20~\gauss$, but this value refers to a spatial resolution corresponding to the grid scale of the MHD simulations, a few $10~\km$.
Therefore, unless there are no magnetic structures below $0.5\arcsec$ in size, the value found by
	\cite{2005A&A...436L..27K}
is not consistent with the same value found by
	\cite{2003ApJ...582L..55D,
	2003A&A...407..741D}.
From high resolution observations obtained by the Hinode Solar Optical Telescope
	\citep[SOT,][]{2008SoPh..249..167T,
	2008SoPh..249..197S,
	2008SoPh..249..233I,
	2008SoPh..249..221S}
and at the Swedish Solar Telescope we know that smaller-scaled structures are quite common.

Recently, analysis of Hinode spectropolarimeter (SP) data by
	\cite{2008ApJ...672.1237L}
has yielded $11~\gauss$ for the longitudinal field.
The fact that Hinode data (at the significantly higher and stable resolution of $0.32\arcsec$) reveal only half as much flux as the investigation of
	\cite{2003ApJ...582L..55D,
	2003A&A...407..741D}
suggests that fluctuations due to seeing may have affected the data of
	\cite{2003ApJ...582L..55D,
	2003A&A...407..741D}
in a way that it increased the strength of the $V$ signal, which is quite conceivable (cross-talk from brightness or velocity into Stokes $V$ can happen quite easily).

An important question is how much flux could be hidden below the spatial scales that can be resolved?
Such a hidden field (i.e., field not visible in magnetograms) is generally referred to as turbulent field, since in order to be invisible the two magnetic polarities must be mixed at scales below the spatial resolution element.
Although very often no clear distinction is made to internetwork fields (which often also show a nearly random distribution of opposite polarities), we could consider internetwork fields as the large-scale and hence roughly resolved parts of the ``turbulent'' field, some fraction of which remains unresolved.
However, given our current knowledge, we cannot rule out that the latter is physically different from the internetwork fields in some important aspect.
	\cite{1959ApJ...129..375U}
was the first to look for an unresolved ``turbulent'' field (that to first order was expected to be isotropic).
Using differential line broadening he was able to set an upper limit of $300~\gauss$ on such a field.
	\cite{1977A&A....59..367S}
and later Stenflo (private communication) greatly improved the sensitivity of the technique by extending the investigation to hundreds of spectral lines (all the unblended \FeI\ lines in the visible solar spectrum), resulting in an upper limit of $100~\gauss$ for the field outside the network, which includes the area-weighted contribution of the internetwork field and of any turbulent field.

Early work on the determination of unresolved magnetic flux using the Zeeman effect was also carried out by
	\cite{1987SoPh..114....1S},
who analyzed Stokes $I$, $Q$, and $V$ profiles and set limits on a combination of magnetic field inclination and field strength.
	\cite{1979ApJ...229..387T}
used high spatial resolution observations to circumvent the problem of cancellation of Stokes $V$ by opposite polarity fields.
They found that a possible turbulent field cannot exceed $100~\gauss$ at spatial scales accessible to observations with a spatial resolution of $0.5\arcsec$.

From the Hanle depolarization of the resonant polarization of lines formed in the quiet sun's photosphere (mainly from the \SrI\ line at $460.7~\nm$, but also from molecular lines), a turbulent magnetic field in the range of roughly $10$--$60~\gauss$ has been inferred
	\citep{1982SoPh...80..209S,
	1995A&A...298..289F,
	2001A&A...378..627F,
	1998A&A...329..319S,
	2004A&A...417..775B,
	2004Natur.430..326T,
	2005A&A...432..295B,
	2006A&A...458..625B,
	2006A&A...457.1047D}.

With time, the investigations have increased in sophistication, now including multi-dimensional polarized radiative transfer and atmospheres produced by 3D ra\-di\-a\-tion-hy\-dro\-dy\-na\-mic simulations.
In general, a field of this average strength covering the whole quiet sun harbors less magnetic energy than the field in the network.
	\cite{2005A&A...438..727S}
has argued, however, that (under certain assumptions) the measurements made in the \SrI\ line actually imply that more than half of the sun's surface is covered by fields stronger than $60~\gauss$, even if the measurements give average field-strength values below $60~\gauss$.

	\cite{2004Natur.430..326T}
also favor a higher energy density in the internetwork than in the network field (deduced from observations obtained at IRSOL by
	\citealp{1997A&A...322..985S}).
They adopt an exponential probability distribution function (PDF) for the field strength, as derived from MHD simulations.
For a single PDF of the magnetic field, they find an e-folding width of $130~\gauss$ (deduced from the same observations as give a $60~\gauss$ average field).
Finally, they introduced different PDFs of the field in granules and intergranular lanes, with $B_0=15~\gauss$ in the former structures and with $B_0=450~\gauss$ in the latter, in order to simultaneously satisfy \SrI\ (atomic) and \CC\ (molecular) lines.
The energy density in the turbulent field in this scenario is larger than in the network.
Further work on this topic, e.g., which tests the assumptions made by
	\cite{2004Natur.430..326T},
would be of considerable interest.

\subsection{Magnetic field strength of quiet sun fields}\label{sec:mfsoqsf}

One question that has led to a partly heated debate over the last decade has been whether the magnetic fields in the internetwork quiet sun are intrinsically weak or strong.
The magnetic field in the network has long been known to have an intrinsic strength on the order of a kilogauss
	\citep[e.g.,][etc.]{1973SoPh...32...41S,
	1978A&A....69..279W,
	1984A&A...140..185S,
	1985SoPh...95...99S,
	1987A&A...173..167S,
	1987A&A...188..183S,
	1992ApJ...391..832R,
	1992ApJ...390L.103R,
	1992A&A...263..323R,
	1996A&A...315..610G},
although a few advocates of weak fields, even in plage and the network remained.
Thus,
	\cite{1989ApJ...340..571Z}
argued that the highly Zeeman sensitive \MgI\ lines at $12.3~\mum$ only show weak fields, so that there are no strong fields in the network or in plages (except in occasional micro-pores).
However, detailed radiative transfer modeling of these lines by
	\cite{1995A&A...293..240B}
has shown that they are formed just below the temperature minimum in plage.
At this height, due to pressure balance with the surrounding gas, the field, which in the middle and lower photosphere is well over a kilogauss, has dropped to only a few hundred gauss.
The observations of
	\cite{1989ApJ...340..571Z}
actually provided confirmation of the simple model of slender flux tubes
	\citep[e.g.,][]{1976SoPh...50..269S},
if extended to take into account the merging of neighboring features
	\citep{1986A&A...154..231P,1986A&A...170..126S}.

More recently, the debate on the intrinsic strength of quiet sun fields has been rekindled, but now concentrating on the internetwork fields.
The intrinsic field strength is much more difficult to measure accurately than the magnetic flux per feature, since the Zeeman splitting often gives a non-unique result, except for kilogauss fields that fill a sufficiently large part of the aperture.
Here, measurements in the infrared have an advantage, since the ratio of Zeeman splitting to Doppler width scales roughly linearly with the wavelength.
It is therefore not so surprising that intrinsically weak fields in the lower photospheric layers were initially observed in the infrared at
$1.56~\mum$
	\citep[e.g.,][]{1992A&A...263..323R}.
It also explains why studies of the strength of internetwork fields that employ infrared data (all have used the Zeeman sensitive line pair at $1.56~\mum$) give consistent results: the field strength of most internetwork features lies below roughly $600~\gauss$
	(\citealp{1995ApJ...446..421L,
	1996A&A...310L..33S,
	2003A&A...408.1115K,
	2005A&A...442.1059K};
see also
	\citealp{1999ApJ...514..448L,
	2007A&A...469L..39M}).
These field strengths are partly consistent with equipartition between magnetic energy density and convective energy density, although they also provide evidence for a partial convective collapse
	\citep{1996A&A...310L..33S}.

Observations of spectral lines in the visible have given rather varied intrinsic strengths of internetwork fields.
Partly the results depend on the employed spectral lines, but they can also differ between studies using the same set of lines.
An initial investigation by
	\cite{1994A&A...286..626K}
employing Stokes $V$ measurements of \FeI\ $525.02~\nm$ and \FeI\ $524.71~\nm$ could not determine the true field strength, but provided evidence for a field strength below a kilogauss.
Interest in these lines has been dormant until very recently when
	\cite{2007ApJ...659.1726K}
and
	\cite{2008ApJ...674..596S}
have studied them in comparison with the more widely used $630.25$ and $630.15~\nm$ line pair as well as the $1.56~\mum$ lines.

Most widely used have been the \FeI\ line pair at $630.25$ and $630.15~\nm$ which have been observed by, e.g., the Advanced Stokes Polarimeter (ASP,
	\citealp{elmore1992})
and now the Hinode SP.
Magnetograms in these lines, recorded with the G\"ottingen Fabry-Perot at the VTT, have been investigated by
	\cite{2003ApJ...582L..55D,2003A&A...407..741D}.
They found magnetic flux throughout the quiet sun (covering $45\%$ of the surface area, typically located in the intergranular lanes).
In addition, this flux, which corresponds to most of the flux in the internetwork, was found to be in the form of kilogauss fields.
Spectroscopic investigations employing ASP data by
	\cite{2004ApJ...613..600L}
did not reproduce the preponderance of strong fields, while the analysis of
	\cite{2004ApJ...616..587S}
indicated a mixture of strong and weak fields
	\citep[cf.][]{2004ApJ...616..587S}.
	\cite{2003ApJ...597L.177S},
from a comparison of visible and infrared lines, also found a mixture of field strengths, although in their case the visible lines gave strong fields, while the infrared lines indicated weak ones (see further below for a more detailed discussion of this result).

Most recently, these lines, as recorded by Hinode/SP, have been analyzed by
	\cite{2007ApJ...670L..61O,
	2007PASJ...59S.837O}.
In contrast to earlier authors, they obtained weak fields with strengths in the range of equipartition with the convection.
Note that in contrast to, e.g.,
	\cite{2003ApJ...582L..55D,
	2003A&A...407..741D},
they inverted the full Stokes vector.

Finally, a Zeeman-effect based diagnostic has been developed by
	\cite{2002ApJ...580..519L,
	2006A&A...454..663L}.
It makes use of the change in line profile shape introduced by hyperfine structure in \MnI\ lines in the visible part of the spectrum ($553~\nm$) as the field strength increases.
Applying this diagnostic to measurements of Stokes $I$ and $V$ in the internetwork they obtain mainly hectogauss fields (which cover the majority of the area and contain the majority of the flux), although they do not give precise numbers regarding the field strength.
This result is confirmed by
	\cite{2007ApJ...659..829A}
employing a \MnI\ line in the infrared (at $1.5262~\mum$).
	\cite{2008ApJ...675..906S},
however, argue that in a MISMA-like atmosphere (Micro-Structured Magnetic Atmosphere, an approximation to describe the influence on the Stokes profiles of an atmosphere with the magnetic field structured at a very small, optically thin scale;
	\citealp{1996ApJ...466..537S})
the \MnI\ line at $553.8~\nm$ will indicate weak fields even if more than $50\%$ of the magnetic flux is in the form of kilogauss fields.

The difference between the results obtained with the infrared $1.56~\mum$ and those from the visible $630.2~\nm$ lines has fueled the aforementioned debate on the true field strengths of internetwork fields.
It has also led the groups using either one of these diagnostics to comment on the shortcomings of the other.
For example, it has been argued that the visible lines miss much of the weak fields, since for incomplete Zeeman splitting
(which is the case for these lines for sub-kilogauss fields)
the signal in a given pixel is proportional to the magnetic flux in that pixel.
Since the internetwork fields are associated with very small fluxes per pixel, these lines could miss a considerable portion of it.
Also, because intrinsically weak fields change the shapes of the $I$ and $V$ profiles only in subtle ways, the deduced values are susceptible to noise or systematic errors.
Conversely, it has been argued by
	\cite{2003ApJ...593..581S}
that the infrared lines, by dint of their large Zeeman sensitivity, give too much weight to the weak fields.
For these lines, the amplitude of the Stokes $V$ signal is proportional to the fractional area covered by the field (the magnetic filling factor) rather than to the amount of magnetic flux in the pixel.
Therefore, fields with a low strength give proportionately stronger signals (for, e.g., an equal amount of magnetic flux in intrinsically weak and intrinsically strong fields).
	\cite{2003ApJ...593..581S}
argue that the rapid drop of the field strength with height (due to pressure balance)
compounds this effect: since a spectral line is formed over a range of heights, this gradient of the field spreads the signal in the wavelength direction.
Since the intrinsically strong fields are associated with the largest vertical field-strength gradients, the smearing in the wavelength direction is largest for such fields, making the Stokes amplitudes small and possibly hidden in the noise.
Consequently, they argue, the infrared lines are missing much of the flux in the strong fields.

\begin{figure}[tbp]
\begin{center}
\includegraphics[width=\textwidth]{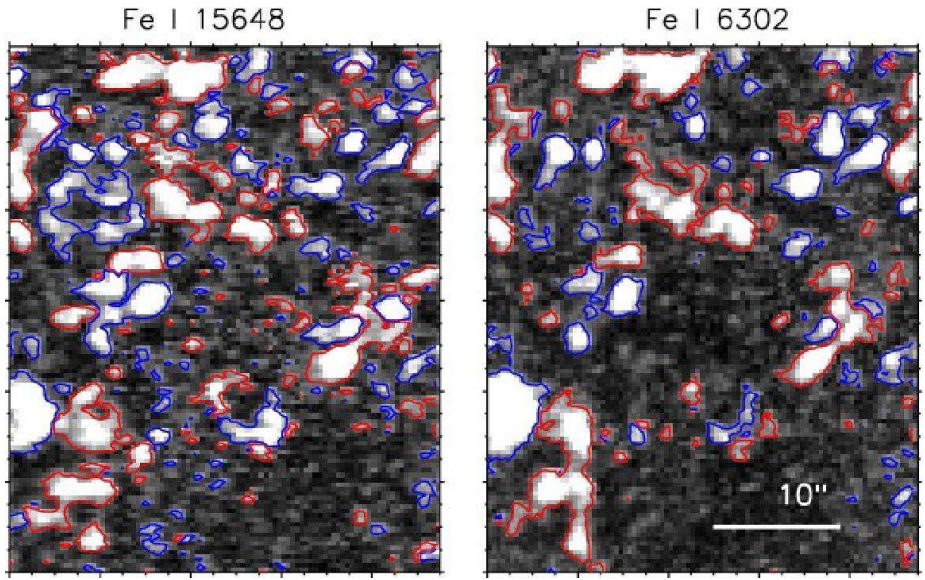}
\caption{Magnetograms in the infrared at $1.56~\mum$ (left panel) and in the visible at $630.2~\nm$ (right panel) obtained simultaneously and co-spatially with the VTT on Tenerife.
Greater brightness indicates larger amounts of magnetic flux per pixel.
Opposite magnetic polarities are bounded by red and by blue lines,
respectively.
Adapted from
	\cite{2005A&A...436L..27K}.}
\label{IRvis_khomenko}
\end{center}
\end{figure}

A comparison of the results obtained from the infrared and the visible lines of a simultaneously observed patch of quiet sun might be a way of deciding between the different diagnostics and associated points of view.
Such a comparison was first carried out by
	\cite{2003ApJ...593..581S},
who found that the Stokes maps in the two wavelength ranges looked quite different.
In particular, they noted that the visible and infrared lines displayed opposite polarities in 25\% of the pixels, which was a remarkably high proportion.
If correct, this would indeed support the view that the infrared and visible lines were sampling rather different components of the internetwork field.
The main drawback with this investigation was that data from different telescopes had been used, so that the seeing quality of the two data sets was not comparable, making their comparison less straightforward
	\citep{1987ApOpt..26.3838L}.
In a later analysis,
	\cite{2005A&A...436L..27K}
compared visible and infrared lines observed with the same telescope under identical seeing conditions and obtained a more similar distribution of polarities and fluxes from both wavelength ranges.
The magnetograms obtained in both wavelength ranges are shown in Fig.~\ref{IRvis_khomenko}.
Remaining differences between the two images are due to the larger sensitivity of the infrared line to weak fields and to the remaining unavoidable differences in seeing (which possesses a dependence on
$\mathrm{\lambda}$).

\begin{figure}[tbp]
\begin{center}
\includegraphics[width=0.48\linewidth]{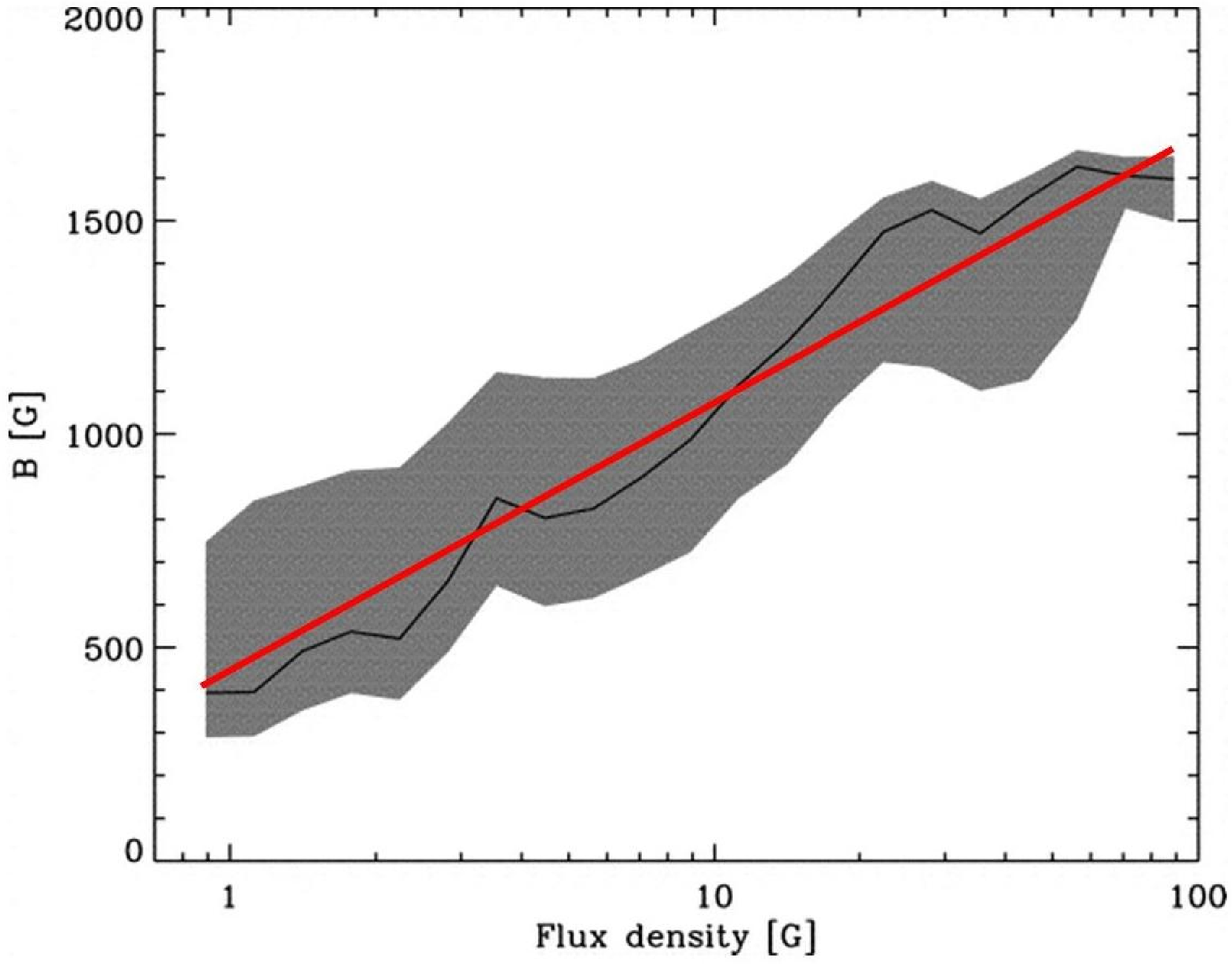}
\hfill
\includegraphics[width=0.48\linewidth]{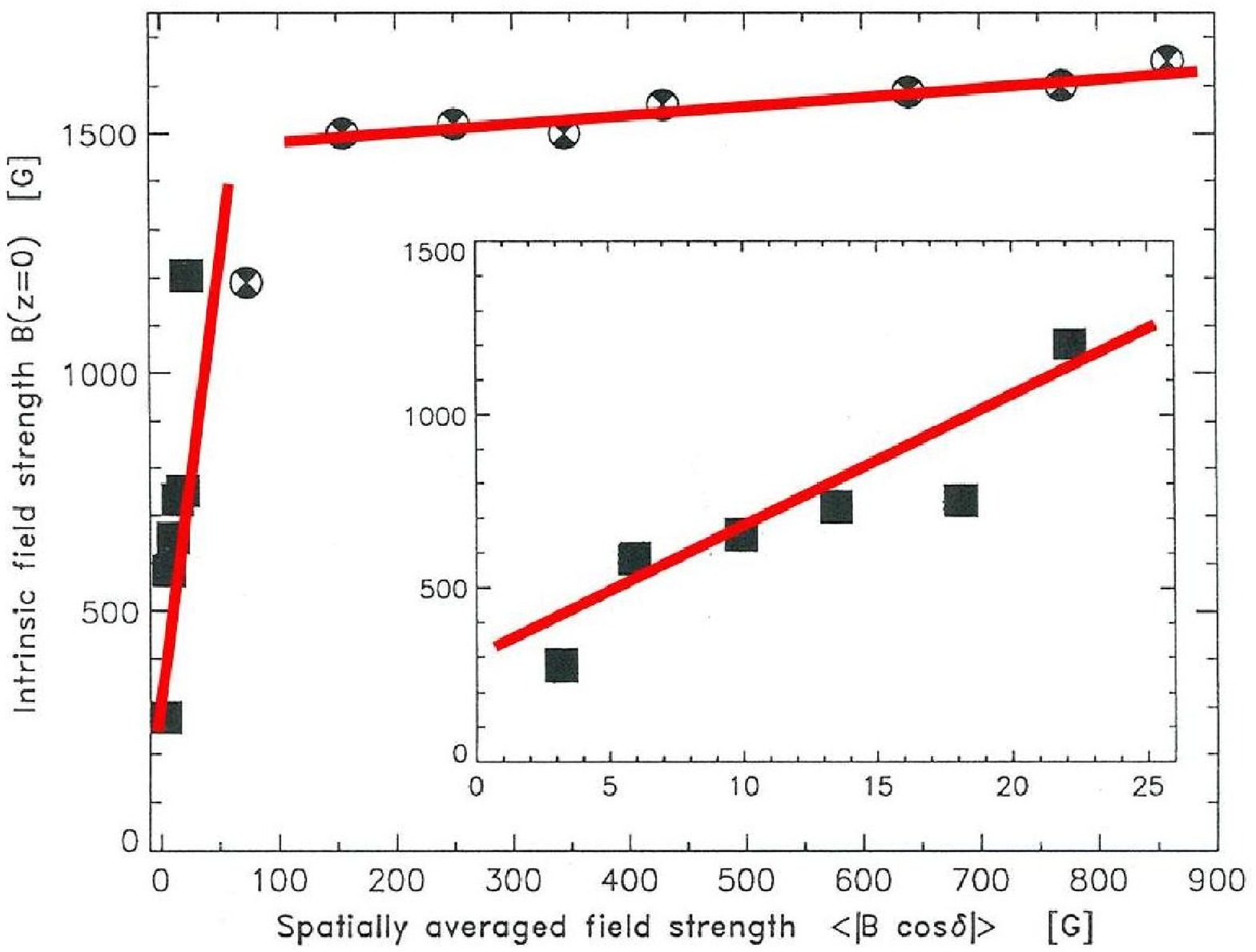}
\caption{%
Left panel: field strengths retrieved from a combination of $1.56~\mum$ and $630.2~\nm$ co-spatial observations obtained nearly simultaneously.
Adapted from
	\cite{2006ApJ...646.1421D}.
Right panel: same, but from an investigation of the $1.56~\mum$ lines alone.
The results of observations covering both the quiet sun and active regions are  displayed in the larger frame.
In the inset only the results for the quiet sun are shown (each point is a value binned over numerous individual data values).
The left red line in the main frame is drawn considering more data than in the
inset (circle at ($\langle B cos\gamma\rangle$, $B) = (70,1200)$) and is not identical with the red line in the
inset.
Adapted from
	\cite{1996A&A...310L..33S}.}
\label{B_vs_avgeB}
\end{center}
\end{figure}

Finally,
	\cite{2006ApJ...646.1421D}
inverted a set of combined infrared and visible spectra using a 3-component model, which allowed a field-free component to co-exist with two different magnetic components.
They obtained a mixture of strong and weak fields, with a clear relationship between magnetic field strength and magnetic flux in the sense that the larger the magnetic flux in a pixel, the stronger the field (left panel of Fig.~\ref{B_vs_avgeB}).
This result is similar to that found by
	\cite{1996A&A...310L..33S},
based purely on $1.56~\mum$ spectropolarimetry, shown in the right panel of Fig.~\ref{B_vs_avgeB}.
An increase of field strength with magnetic flux of the feature is in agreement with predictions of the efficiency of the convective collapse mechanism that leads to the formation of the intense flux tube
	\citep{1986Natur.322..156V}.
As in their earlier papers,
	\cite{2006ApJ...646.1421D}
argue that most of the flux and of the magnetic energy is in the kilogauss fields.

Many of the investigations discussed so far have been based on Stokes $I$ and $V$ profiles only.
This can be explained partly by instrumental constraints, partly by the fact that the Stokes $Q$ and $U$ profiles scale as $B^2$, while Stokes $V$ scales proportionally to $B$.
For a relatively weak Zeeman splitting (typical of visible lines in the quiet sun) this implies that $Q$ and $U$ are much weaker than $V$.

The difficulty of measuring the field strength reliably from just $I$ and $V$ of a visible line pair, in particular from \FeI\ $630.2$ and $630.1~\nm$, has been demonstrated by
	\cite{2006A&A...456.1159M}.
They fit a set of these line profiles two times, once starting from a strong-$B$ initial guess, once from a weak-$B$ initial guess.
The final result depended strongly on the initial guess, although the fits to the profiles were equally good.
The differences in the Stokes profiles produced by the different field strengths were completely compensated by slightly different temperatures and turbulence velocity values returned by the inversion code.
Another test was carried out by
	\cite{2007ApJ...659.1726K}.
They used the output atmospheres from the 3D radiation-MHD simulations of
	\cite{2005A&A...429..335V}
to test a number of diagnostics of the field strength.
According to their analysis the most reliable of the tested diagnostics is the $1.56~\mum$ line pair, the least reliable the $630.2$/$630.1~\nm$ line pair.
These exercises have demonstrated just how difficult it is to obtain reliable $B$ values from this latter line pair, in particular if only Stokes $I$ and $V$ are available.
Consequently, results obtained from such data have to be interpreted with caution.

The great advantage of also having a linear polarization profile available is that the shape of the $Q$ and $U$ profiles changes with the field strength, in the sense that the ratio of the strength of the $\pi$-component to the $\sigma$-components depends on $B$, providing further (although not in themselves unique) constraints on the field strength, as demonstrated by,
e.g.,
	\cite{1987A&A...188..183S}.

More recently, the advent of Hinode has opened up new possibilities, by providing $I$, $Q$, $U$, and $V$ spectra of \FeI\ $630.2$ and $630.1~\nm$ at a constant high spatial resolution corresponding to approximately $0.3\arcsec$.
Recent inversions by
	\cite{2007ApJ...670L..61O,2007PASJ...59S.837O}
indicate that the Hinode data give mainly weak fields (hectogauss), possibly because of the additional constraints provided by the linear polarization signals (only pixels with profiles lying above a given threshold in Stokes $Q$, $U$, and $V$ are inverted).

\subsection{Horizontal fields in the internetwork}\label{sec:hfiti}

Evidence for horizontal fields in the internetwork can be noted already in data published by
	\cite{1988SoPh..117..243M}:
in these magnetograms internetwork fields are visible from the center of the solar disk right to the limb, suggesting the presence of both vertical and horizontal fields.
With considerable foresight, Martin interpreted these measurements as possibly due to the presence of low-lying loops in the internetwork.

	\cite{1996ApJ...460.1019L}
found arcsecond scale, short-lived horizontal fields (lifetimes of minutes) in the internetwork.
The size scale was determined by their spatial resolution.
	\cite{1998A&A...331..771M}
considered the center-to-limb variation of the Stokes $V$ amplitude of the $g=3$ line at $1.56~\mum$ from which they concluded that the quiet sun field is composed mainly of intrinsically weak, nearly isotropically distributed fields, in addition to strong, nearly vertical fields.
	\cite{2008A&A...479..229M}
also found evidence for a more or less isotropic distribution of the internetwork field
(and little change in the field strength probability distribution function)
from the center-to-limb variation of the polarization signal in the quiet sun,
in agreement with
	\cite{1988SoPh..117..243M}
and
	\cite{1998A&A...331..771M}.
With the very sensitive SOLIS instrument on Kitt Peak,
	\cite{2007ApJ...659L.177H}
deduced a ``seething'' horizontal field throughout the internetwork.
This field of typically $1$--$2~\gauss$ at the spatial resolution of SOLIS of $2.5$--$5\arcsec$ changed within minutes.
Further evidence for horizontal fields has been provided by Hinode:
	\cite{2007ApJ...670L..61O,2007PASJ...59S.837O}
inverted Stokes spectra to obtain a peak in the distribution of inclination angles of internetwork fields at $90^\circ$, which corresponds to horizontal fields.
This interesting result may partly be an artifact of the higher sensitivity to noise of Stokes $Q$ and $U$ due to their weakness, unless fields are intrinsically strong.
Finally,
	\cite{2008ApJ...672.1237L}
obtained 5~times more flux in horizontal fields than in the vertical fields in the internetwork (to be more specific: they found that the spatially averaged strength of the horizontal field is 5 times larger than of the vertical field; a precise determination of the flux for horizontal fields is rather difficult from Stokes parameters).
With a strength of $50$--$60~\gauss$, it is comparable to the values obtained by the Hanle effect (see Sect.~\ref{sec:urmf}).

\begin{figure}[tbp]
\begin{center}
\includegraphics[width=\textwidth]{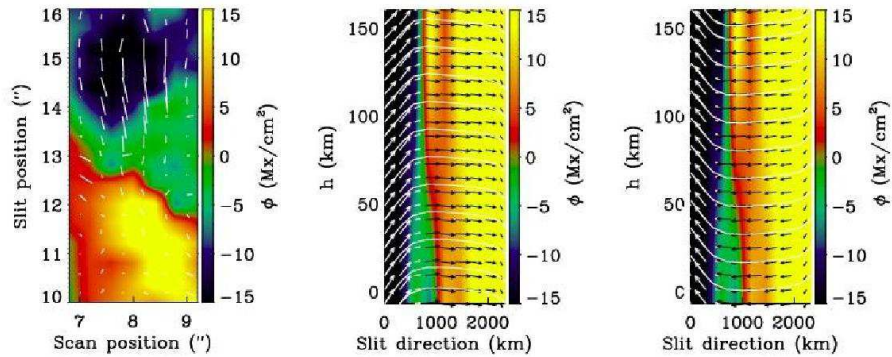}
\caption{Reconstructed loop in the internetwork.
Left panel: vertical magnetic flux density in a small region of the total scan (flux density is indicated by the color) at height zero
(average solar surface).
The azimuthal direction of the field is overplotted.
Central panel: the vertical dependence of the magnetic vector along a cut going from the upper part of the left frame to its lower part at the scan position marked $8\arcsec$.
The colors mark the magnetic flux density, while the direction of the magnetic vector is indicated by the arrows.
The white lines are smoothed curves joining the arrows and outlining the loops.
The right panel is the same, but now for the other solution allowed by the $180^\circ$ ambiguity.}
\label{IN_loops}
\end{center}
\end{figure}

As the evidence for nearly horizontal internetwork fields increases, one question that comes to the fore is: what is the structure of these internetwork fields?
From observations at $1.56~\mum$,
	\cite{2007A&A...469L..39M}
concluded that at least some of the internetwork elements are (parts of)
low-lying loop-like structures.
The loops were reconstructed in a way similar to the technique applied by
	\cite{2003Natur.425..692S},
although the $180^\circ$ ambiguity inherent in the Zeeman-effect did not allow
	\cite{2007A&A...469L..39M}
to distinguish between small $\mathrm{\Omega}$ loops and U-loops at a granular scale.
An example of a loop reconstructed by
	\cite{2007A&A...469L..39M}
is shown in Fig.~\ref{IN_loops}.
These loops may correspond to the small-scale emerging loops observed by
	\cite{2007ApJ...666L.137C}
in the quiet sun and by
	\cite{2008A&A...481L..25I}
in active region plage.
These small loops carry a flux of approximately $10^{17}~\mx$ each and are found to emerge in granules.

\subsection{Source of internetwork fields}

How could such a magnetic structure be explained? There are different possibilities.

\emph{1.~Emergence of fields generated in deeper layers} (e.g., by a deep convection-zone or an overshoot-layer dynamo).
This could be the extension of ephemeral active-region fields (studied by
	\citealp{1973SoPh...32..389H,
	1975SoPh...40...87H,
	HarveyPhDT,
	2001ApJ...555..448H})
to still smaller scales.
Note that there is a power-law distribution of flux in bipolar regions (following an inverse square law) from large active regions down to small ephemeral regions.
The cutoff at the small scales is consistent with a lack of resolution and/or sensitivity.
Whereas the large active regions have a strong tendency towards an E--W orientation following Hale's polarity law, increasingly smaller bipoles have increasingly weaker preferred orientations.
Any lack of orientation of the smallest emerging bipoles therefore does not automatically rule out this scenario, since there is no abrupt transition,
but rather a very gradual decrease of the level of orientation with decreasing area or magnetic flux.

\emph{2.~Flux recycling after decay of active-regions and ephemeral active regions.} The magnetic flux from a decaying region very likely partly gets dragged down by convection and can emerge again at another point on the solar surface.
Such an effect has been identified in MHD simulations carried out by
	\cite{2001ASPC..236..363P},
suggesting that such recycling does take place.
The work of
	\cite{2005A&A...441.1183D},
see below, also provides evidence that either mechanism~1 or~2 (or some combination of both) is acting as the source of some of the flux in the internetwork (see Sect.~\ref{sec:bpme}).

\emph{3.~Flux produced at or very close to the solar surface by a truly local dynamo.}
The first numerical experiments that sustained a local dynamo in a convective medium similar to the solar interior were carried out by
	\cite{1999ApJ...515L..39C}.
The most realistic simulation of a local (solar surface) dynamo to date has been performed by
	\cite{2007A&A...465L..43V},
who considered also a proper 3D radiative transfer etc.
to simulate the conditions in the layers close to the solar surface.
Starting from a very low value, the magnetic energy within the simulation box increases exponentially with time, before it saturates.
The saturation value depends on the magnetic Reynolds number $R_m$ of the simulation, being higher for larger $R_m$.
For $R_m=2600$ the simulations give an average, unsigned vertical field of approximately $35~\gauss$, which lies within the range of values found from the Hanle effect.

\begin{figure}[tbp]
\begin{center}
\includegraphics[width=\textwidth]{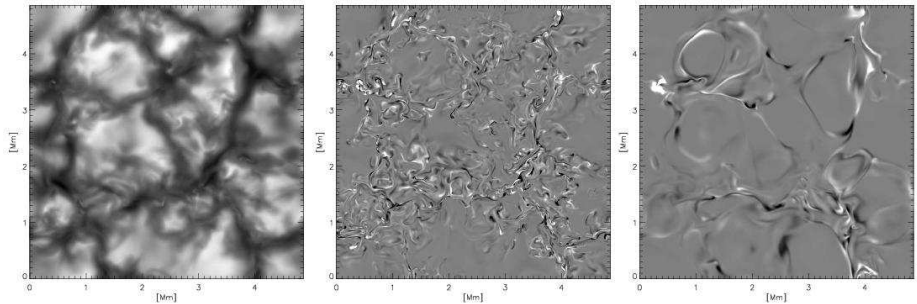}
\caption{Snapshot from a dynamo simulation run taken about 5~hours after introducing the seed field.
The vertically emerging bolometric intensity (brightness, left panel) reveals a normal solar granulation pattern.
The other panels show the vertical component of the magnetic field on two surfaces of constant (Rosseland) optical depth, $\tau_R$.
Near the visible surface (middle panel, $\tau_R = 1$, gray scale saturating at $\pm250~\gauss$), the magnetic field shows an intricate small-scale pattern with rapid polarity changes and an unsigned average flux density of $25.1~\gauss$.
About $300~\km$ higher, at the surface $\tau_R = 0.01$ (right panel, gray scale saturating at $\pm50~\gauss$), the unsigned average flux density has decreased to $3.2~\gauss$ and the field distribution has become considerably smoother, roughly outlining the network of intergranular downflow lanes (darker areas on the left panel).
Figure taken from
	\cite{2007A&A...465L..43V}
by permission.}
\label{turb_dynamo}
\end{center}
\end{figure}

The field produced by such a simulation is structured on very small (subgranular) scales with strongly mixed opposite polarities, as can be seen from Fig.~\ref{turb_dynamo}.
It is also largely horizontal.
It is basically composed of short, flat loops that are concentrated in intergranular lanes and generally have both their foot points within a single intergranular lane.
Note that the simulations carried out so far do not allow any flux to be advected into the box (which may be the reason why relatively few larger-scale magnetic structures are visible).
Note also that changes in $R_m$ should have an influence on the magnitude of the produced magnetic field and energy, but not on its distribution, so that the shape of the PDF of the field strength and of the magnetic orientation should remain independent of $R_m$.

Any difference between the observed and simulated distribution of the flux may be telling us something about other effects besides a purely local dynamo acting to produce the observed field.
Therefore, it is heartening that
	\cite{2008A&A...481L...5S}
obtain a ratio between horizontal and vertical field that is close to the value found by
	\cite{2008ApJ...672.1237L}
from Hinode SP data.
One difference between the two is that
	\cite{2008ApJ...672.1237L}
found most of their horizontal flux regions at the edges of granules, while simulations place the flux clearly in the intergranular lanes.
A part of this difference may be due to the limited depth of the computation box.

Quite generally, there is an observed relationship between the weak quiet-sun fields and convective features.
Best known is that the strong fields found in the network are located at the boundaries of supergranules.
On a smaller scale,
	\cite{1999ApJ...514..448L}
find a weak field whose distribution is moulded by the granulation.
The field also changes over a granular life-time (consequently they called this component of the field a granular field).
	\cite{2003A&A...408.1115K}
find a preponderance of weak fields in the intergranular lanes, while
	\cite{2004ApJ...611.1139S}
find that the field strength depends on the location of the field relative to the granule in a non-trivial manner.
Arguments against the origin of at least the stronger internetwork flux from a local dynamo have been given by
	\cite{2005A&A...441.1183D}
on the basis of the fact that this part of the flux is seen to be distributed on a mesogranular scale and displays a lifetime well in excess of that of granulation.

Such dependencies may (or may not) provide an indication of the origin of the magnetic flux.
However, they do tell us that the flux must survive without complete cancellation for a sufficiently long time to be dragged to the edge of the particular convective feature it is found to be lying at the boundary of.
In the case of the network this implies a survival time of at least 10~hours, for the mesogranulation roughly an hour or two.

\section{Transient Horizontal Magnetic Field}
\label{sec:thmf}
\subsection{Properties of horizontal magnetic field}

Quiet-sun magnetism essentially consists of vertical flux tubes and horizontal magnetic fields.
The field strength of vertical magnetic fields exceeds the equipartition field strength $B_{e}$ of about $500~\gauss$, determined by $B_{e}=\sqrt{4\pi\,\rho\,v^2}$, where granules with a velocity of $v=2\times10^5~\cms$, and the plasma density $\rho=3\times10^{-7}~\gccm$ at $\tau_{500}=1$ are assumed.
Hinode observations show that convective instability could be a mechanism used to explain the formation of such vertical flux tubes with kilogauss field strength \citep{2008ApJ...677L.145N}: the cooling of a flux tube at equipartition field strength precedes a transient downflow reaching $6~\kms$ and the intensification of the field strength to $2~\kgauss$.
This is not a unique observation, but rather it is a ubiquitous phenomenon in the quiet sun.

The initial discovery of the horizontal magnetic field with ground-based telescopes was summarized in the previous section.
High resolution spectroscopic observations with SOT/SP aboard Hinode have confirmed this finding and extended these studies considerably \citep{2008ApJ...672.1237L, 2007ApJ...666L.137C, 2007ApJ...670L..61O, 2007PASJ...59S.837O, 2008A&A...481L..25I, Ishikawa2008b, tsuneta2008}.
The horizontal magnetic field is highly intermittent in both the temporal and spatial domain: statistical study shows that the average life time of a horizontal field element is 4 min, and their size is smaller than the average size of the granular pattern \citep{Ishikawa2008c}.
Thus, we hereafter call them elements of the ``transient horizontal magnetic field'' (THMF).

\begin{figure}[tbp]
\begin{center}
\includegraphics[width=0.67\textwidth]{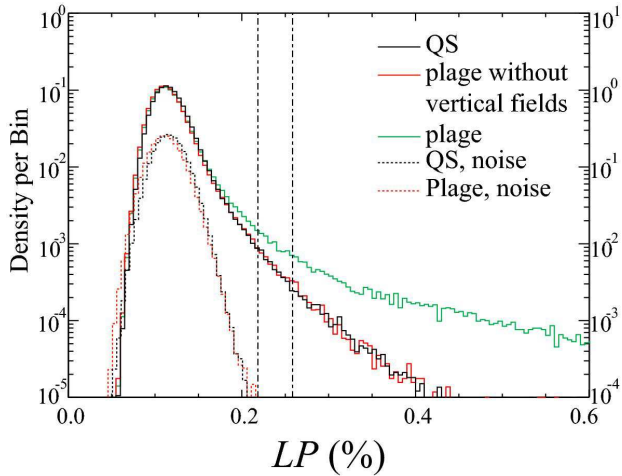}
\caption{Histograms of net linear polarization (LP) for plage and quiet sun.
The dotted lines represent LP noise distributions for both datasets.
The area dominated by vertical magnetic fields is masked in the plage region.
The two vertical dashed-dotted lines indicate LP of 0.22\% and 0.26\% -- the thresholds used in Figure 3.
From \cite{Ishikawa2008b}.}
\label{LPhist}
\end{center}
\end{figure}

Figure~\ref{LPhist} shows the histogram (i.e., PDF) of the degree of linear polarization (LP) for THMF in the quiet sun and a plage region.
These two regions were located near the center of the solar disk.
The degree of linear polarization is proportional to the square of the transverse magnetic field component.
Vertical magnetic concentrations are masked, and thus are not included in the red histogram.
This is a comparison of different areas of the sun with different magnetic properties.
The exact match of these two PDFs indicates that property of the THMF of the quiet sun and active regions is remarkably similar \citep{Ishikawa2008b}.

The magnetic landscape of the polar region is characterized by vertical kilogauss patches with super equipartition field strength, a coherency in polarity, and the ubiquitous weaker transient horizontal fields \citep{tsuneta2008}.
We now know that THMFs are ubiquitous in plage regions, the quiet sun, and the extreme polar region.

The remarkably similar distributions of LP in Fig.~\ref{LPhist} also suggest the same occurrence rates in both the quiet sun and the plage region.
These occurrence rates are extremely high, as discussed by \cite{2008A&A...481L..25I}.
THMFs have lifetimes ranging from one minute to about ten minutes, comparable to the lifetime of granules.
Among 52 events that they examined, 43 horizontal magnetic structures appear inside the granules, and four appear in inter-granular lanes, with the remaining five events ambiguous in position.
Since 52 events are detected in the $2.5\arcsec\times164\arcsec$ observing area during the 40 minutes, a new event appears every 46 seconds in the same observing region.
The turnover time of the granules is approximately 1000~s, with a velocity of $2~\kms$ and with a depth comparable to the horizontal scale of granules.
There are approximately 182~granules in the observing area, assuming that the size of the granules is $1.5\arcsec\times1.5\arcsec$.
84\% of these granules are not associated with stable strong vertical magnetic fields, and we use this smaller sample for estimating the frequency of events.
If every granule were to have an embedded horizontal magnetic field structure, the horizontal field would have appeared at the surface every $6.6~\mathrm{s}$ ($\sim1000~\mathrm{s}/152~\mathrm{granules}$) in the observing area.
This shows that more than approximately 10\% of the granules have embedded horizontal fields, suggesting a relatively common occurrence of THMFs \citep{2008A&A...481L..25I}.

\begin{figure}[tbp]
\begin{center}
\includegraphics[angle=90,width=0.9\textwidth]{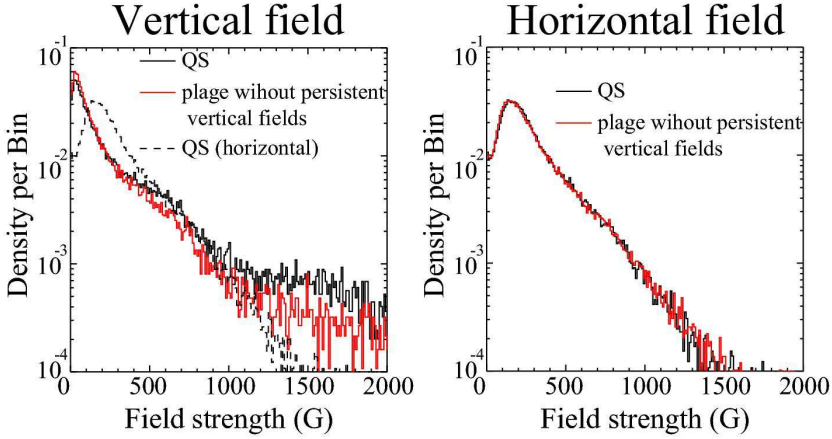}
\caption{
PDFs of the magnetic field strength of vertical fields (left panel) and horizontal fields (right panel).
Black and red lines are used for the quiet sun and the plage region, respectively.
Apparent concentrations of the vertical magnetic fields are masked to obtain the PDF of the plage region.
The black dashed line in the left panel shows the PDF for horizontal fields in the plage region, and is the same as the black solid line in right panel.
Vertical fields refer to magnetic fields with inclination smaller than $20^\circ$ or larger than $160^\circ$, and horizontal fields refer to magnetic fields with inclination larger than $70^\circ$ and smaller than $110^\circ$.
From \cite{Ishikawa2008b}.}
\label{VH_PDF}
\end{center}
\end{figure}

Figure~\ref{VH_PDF} shows that PDFs of the intrinsic magnetic field strength for the quiet sun and the plage region are again almost identical, and the PDF of the extreme polar region (Fig.~\ref{fig:sphist}) is similar to those of the quiet sun and the plage region.
This remarkable similarity suggests a common local dynamo process \citep{1999ApJ...515L..39C,2007A&A...465L..43V} taking place all over the sun.

To minimize the influence of noise in the Stokes inversion, we have analyzed only pixels whose polarization signal peaks exceed a given threshold above the noise level $\sigma$.
The noise level was determined in the continuum wavelength range of the profiles.
The fitting is performed for pixels whose $Q$, $U$, or $V$ signals are larger than $4.5$--$5.0\sigma$.
Thus, the peaks in the PDFs at around $150~\gauss$ may be an artifact: the THMFs that we observe are probably the tip of the iceberg due to our limited sensitivity, and there may be weaker but more ubiquitous magnetic fields unresolved by Hinode: the sun's hidden magnetism inferred by, e.g., \cite{2004Natur.430..326T} through Hanle-effect observations (see Sect.~\ref{sec:urmf} for a more complete discussion).

Figure~\ref{VH_PDF} indicates that 93\% of horizontal magnetic fields have field strengths smaller than $700~\gauss$, and 98\% smaller than $1~\kgauss$ for both regions.
A magnetic field strength of $700~\gauss$ corresponds to the typical equipartition field strength just below the level of granules at a depth of $500~\km$, where the density is $\sim10^{-6}~\gccm$ and the velocity is $2~\kms$.
Thus, the majority of horizontal fields have field strengths smaller than the equipartition field strength for average granular flows.

\begin{figure}[tbp]
\begin{center}
\includegraphics[angle=90,width=0.67\textwidth]{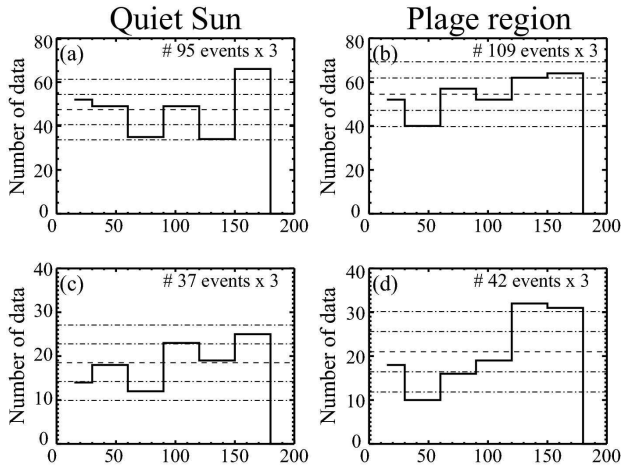}
\caption{The histograms of the field azimuth angles for THMFs.
Three pixels with highest LP are taken from individual events, and number distributions of the azimuth angles of horizontal magnetic field for these pixels are plotted.
The azimuth angle $0^\circ$ is to the west, $90^\circ$ to the north, and $180^\circ$ to the east.
Panels~a and~b: histograms of azimuth angle for 95~events in the quiet sun and 109~events in the plage region which have LP higher than 0.22\%.
Panels~c and~d: histograms of azimuth angle for 37~events in the quiet sun and 42~events in the plage region which have LP higher than 0.26\%.
The dashed lines indicate the case for a uniform distribution.
Dashed-dotted lines closer to the dashed line show $\pm \sigma$, statistical deviation, and two other dotted lines show $\pm$ 2$\sigma$.
From \cite{Ishikawa2008b}.
}
\label{azimuthhist}
\end{center}
\end{figure}

Figure~\ref{azimuthhist} panels~a and~b show the magnetic field azimuth of THMFs for events with LP greater than $0.22\%$ in the quiet sun and in the plage region discussed above.
We define $\sigma=\sqrt{N}$, where $N$ is a number of average events per $30^\circ$ bin under the assumption of a uniform distribution of the azimuth of the horizontal fields.
There is no statistically significant orientation in either region.
If the LP threshold is $>0.26\%$ (Fig.~\ref{azimuthhist} panels~c and~d), we find a broad peak between $120^\circ$ and $180^\circ$, and a dip between $30^\circ$ and $60^\circ$ that are significant at the $2\sigma$ level in the plage region.
In contrast, these events in the quiet sun still show an azimuth angle distributed within $2\sigma$ of the uniform value.
This peak angle corresponds to the tilt angle of the bipolar plage region.
This indicates that THMFs with higher LP in the plage region appear to be partially related to the global fields of the plage region.

\subsection{THMF and local dynamo process}

The properties of THMFs are summarized to be: (1) an identical or similar PDF of magnetic field strength in the quiet sun, plage regions, and the extreme polar region; (2) ubiquitous occurrence all over the sun including the extreme polar region; (3) a magnetic field strength essentially smaller than the equipartition field strength; and (4) no or weak preferred direction of the magnetic field vector.

The amount of the vertical magnetic flux in the plage in the case presented here is about 8~times larger than that of the quiet sun \citep{Ishikawa2008b}.
If the THMF occurrence rate was in any way directly related to the global vertical fields forming the plage region, then we would expect the occurrence rate in the plage region to be much larger than that of the quiet sun.
The similar occurrence rates we observe suggest that the emergence of the THMFs does not have a direct causal relationship with the vertical magnetic fields in the plage region.
The same THMF occurrence rate, no preferred orientation, and similar field-strength distributions for both regions strongly suggest that a common local process that is not directly influenced by global magnetic fields produces THMFs \citep{Ishikawa2008b}.
As (1) ubiquitous THMFs are receptive to convective motion \citep{2007ApJ...666L.137C,2008A&A...481L..25I}, and (2) the field strength is essentially smaller than the equipartition field strength, a reservoir of THMFs may be located near solar surface, and these magnetic fields are carried to the surface through convective flow.

Such reservoir can be maintained by a local dynamo process due to near-surface convective motion \citep{1999ApJ...515L..39C,2007A&A...465L..43V}.
Indeed, numerical simulations have shown that a local dynamo can generate horizontal magnetic structures in the quiet sun \citep{2007ApJ...665.1469A, 2008A&A...481L...5S}.
Such a local dynamo process could naturally explain the similarity in occurrence rates and field strength PDFs, including the fact that THMFs do not have a preferred orientation.
The similarity in field-strength distribution also indicates that properties of THMFs do not depend on the seed field, e.g., global fields.

Other possibilities for the origin of THMFs include debris from decaying active region, magnetic fields that failed to emerge from the convection region to the photosphere \citep{2001ApJ...549..608M}, and extended weak magnetic fields in the upper convection zone generated by ``explosion'' \citep{1995ApJ...452..894M}.
If the reservoir is maintained by one of these processes, the THMFs would be expected to be affected by global toroidal fields in terms of properties of THMFs described above.
Within the context of these simulations cited above, it may be difficult to explain the observed properties of THMFs such as the similarity in the occurrence rates and magnetic field distributions, and the lack of preferred orientation of THMFs.

A slight preferred orientation of THMFs with higher LP toward the global plage polarity suggests that these THMFs may be influenced by the global plage field.
However, because any strong vertical fields associated with the emergence of these THMFs are not observed \citep{2008A&A...481L..25I}, they are probably not directly created from the vertical magnetic fields forming the plage as suggested by \cite{2008ApJ...679L..57I}.
Thus, even if these THMFs with higher LP are related by the global toroidal system, the relationship would be indirect --- the THMFs with high LP may result from fragmented elements of plage flux tossed about by the convective motions below the photosphere.

The evidence that a local dynamo is playing a significant role for the quiet sun magnetism comes from the Hanle-effect investigation by \cite{2004Natur.430..326T}, who inferred a magnetic energy density which is the order of 20\% of the kinetic energy density produced by the convective motions in the quiet solar photosphere, and showed that the observed scattering polarization signals do not seem to be modulated by the solar cycle.
The papers using observations from Hinode cited here are providing us with multiple new pieces of evidence in favor of a local dynamo process taking place in the convective turbulent outer layer of the sun.

\section{Polar field}
\label{sec:pf}
The sun's polar magnetic fields are thought to be the direct manifestation of the global poloidal fields in the interior, which serve as seed fields for the global dynamo that produces the toroidal fields responsible for active regions and sunspots.
The polar regions are also the source of the fast solar wind.
Although the polar regions are of crucial importance to the dynamo process and acceleration of the fast solar wind, its magnetic properties are poorly known.
Magnetic field measurements in the solar polar regions have long been a challenge: variable seeing combined with the strong intensity gradient and the foreshortening effect at the solar limb greatly increases the systematic noise in ground-based magnetographs.
Nevertheless, pioneering observations have been carried out for the polar regions
	\citep{1991SoPh..132..247T, 1994SoPh..155..243L, 1996SoPh..163..223L, 1997SoPh..175...81H, 2004A&A...425..321O, 2007A&A...474..251B}.
Many polar observations have also been restricted to individual polar faculae within a small field of view, and have not provided us with a global magnetic landscape of the polar region, with the exception of GONG/SOLIS
	\citep{2007ApJ...659L.177H}.
Using SOT on board the Hinode spacecraft, it is possible to investigate the properties of photospheric magnetic field in polar regions with unprecedented spatial resolution, field of view, and polarimetric sensitivity and accuracy in measurements of vector magnetic fields.
Such an analysis has recently been carried out by
	\cite{tsuneta2008}.

\subsection{The polar magnetic landscape}

\begin{figure}[tbp]
\begin{center}
\includegraphics[width=\textwidth]{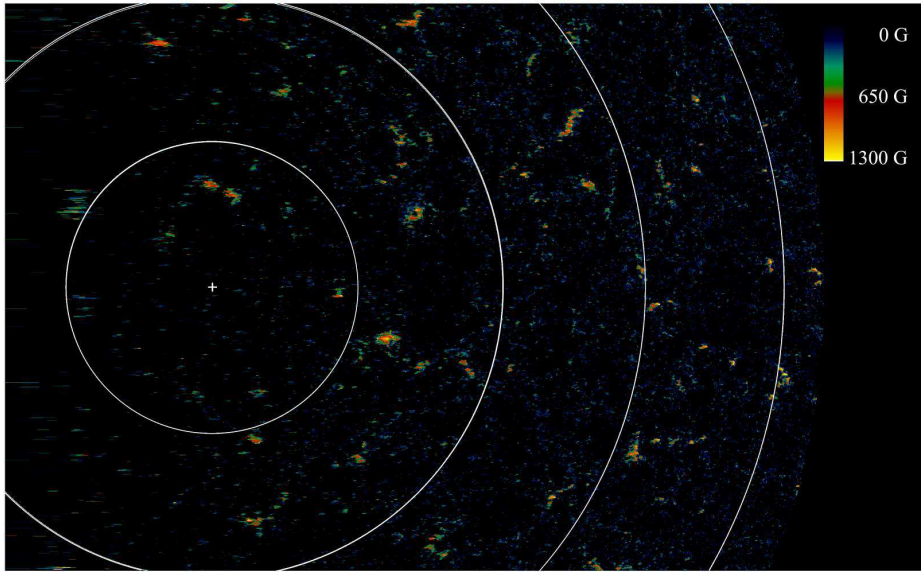}
\caption{View of the magnetic field strength in the south-polar region as seen from above the pole at 12:02:19--14:55:48 on March 16, 2007.
East is up, and north is to the right.
The original field of view of the observation is $328\arcsec$ (east-west) by $164\arcsec$ (north-south).
Latitudinal lines for $85^{\circ}$, $80^{\circ}$, $75^{\circ}$, and $70^{\circ}$ are shown as white circles, while the cross mark indicates the south pole.
The spatial resolution is lost near the extreme limb (i.e., near the left of the figure).
Magnetic field strength is obtained for pixels with a polarization signal exceeding $5\sigma$ above the noise level.
From \cite{tsuneta2008}.}
\label{fig:spb}
\end{center}
\end{figure}

Properties such as field strength, inclination, azimuth, filling factor, etc.\ may be estimated from the line profiles observed by Hinode using inversion codes.
In this case, only pixels whose polarization signal exceeds $5\sigma$ above the noise level were analyzed using an inversion code that assumes a Milne-Eddington atmosphere.

Figure~\ref{fig:spb} is a map of the magnetic field strength as seen from above the south pole.
Such a representation is needed to correctly see the spatial extent and size distribution of the magnetic islands in the polar region.
While many of them are isolated, and some have the form of a chain of islands, complex internal structures are seen inside the individual patches.
Many patchy magnetic islands have very high field strength reaching above $1~\kgauss$.
They are coherently unipolar, and like plage and network fields at lower latitudes
	\citep{1997ApJ...474..810M},
they have magnetic field vertical to the local surface.

Patches show a tendency to be larger in size with increasing latitude.
The size is as large as $5\arcsec\times5\arcsec$ at higher latitudes and $1\arcsec\times1\arcsec$ at lower latitudes.
Degradation in spatial resolution due to the projection effect may contribute to the larger size at high latitude.
Expansion may also be caused because we observe flux tubes higher in the atmosphere close to the limb.
The response function of the spectral lines observed here for a plane-parallel atmosphere viewed obliquely at an angle of $80^{\circ}$ has a peak that is $50$ to $100~\km$ higher than if viewed straight down.

Close to the limb, it is possible to determine the inclination $i$ of the magnetic field vector with respect to the local surface without the usual 180-degree ambiguity of the transverse field components
	\citep{dti04}.
All the large patches have fields that are vertical to the local surface, while the smaller patches tend to be horizontal.
Most of the magnetic structures seen in Fig.~\ref{fig:spb} thus have either vertical or horizontal directions.
These two types do not appear to be spatially correlated.

\begin{figure}[tbp]
\begin{center}
\includegraphics[width=\textwidth]{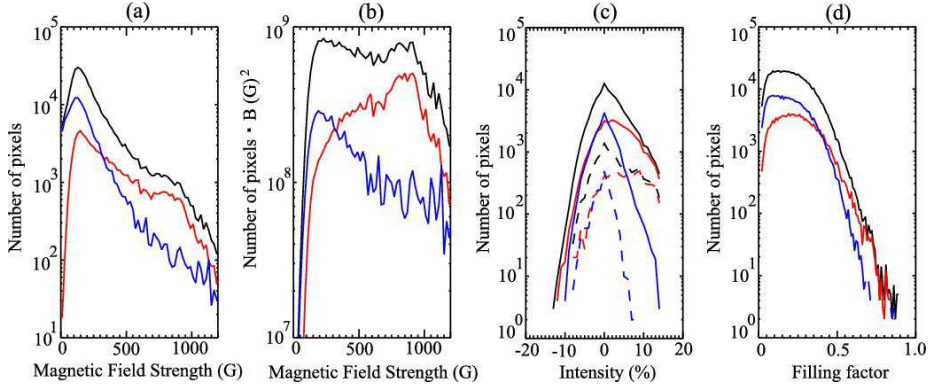}
\caption{Histograms of pixels at latitudes greater than $75^\circ$.
Red lines indicate vertical field, blue horizontal field, and black both.
Panel a: number of pixels as a function of the magnetic field strength (probability distribution function).
Panel b: number of pixels multiplied with $B^2$ as a function of the magnetic field strength.
The bin size of magnetic field strength in panels~a and b is $20~\gauss$.
Panel c: histogram of continuum intensity with magnetic field strength $>300~\gauss$ (solid line) and $>800~\gauss$ (dashed line).
Since continuum intensity rapidly decreases toward the limb, the horizontal axis is the normalized excess continuum level with respect to the continuum level averaged over a $6.4\arcsec$ box.
Panel d: filling factor.
From \cite{tsuneta2008}.}
\label{fig:sphist}
\end{center}
\end{figure}

Magnetic patches of larger spatial extent coincide in position with polar faculae \citep{1994SoPh..155..243L, 2004A&A...425..321O}.
This is confirmed in panel~c of Fig.~\ref{fig:sphist}.
The distribution of local intensity is essentially symmetric around the average intensity for the horizontal fields, while the vertical fields tend to have higher continuum intensities.

Panel~a of Fig.~\ref{fig:sphist} shows the PDF of the magnetic field strength $B$ for latitudes $>75^{\circ}$.
Vertical magnetic fields with inclination $i<25^{\circ}$ dominate the stronger field regime, while horizontal fields with $i>65^{\circ}$ are much more prevalent below $250~\gauss$.
A PDF of the magnetic energy is shown in Fig.~\ref{fig:sphist} panel~b.
This shows that the vertical flux tubes with higher field strength are energetically dominant, while weaker horizontal flux tubes contrastingly carry more energy.

\subsection{Total magnetic flux}

The total vertical magnetic flux in the SOT field of view is $2.2\times10^{21}~\mx$, while the total horizontal flux is $4.0\times10^{21}~\mx$, assuming the filling factor given by the inversion of the data.
The distribution of the filling factors has a broad peak at $f=0.15$ with FWHM range $0.05<f<0.35$ (see Fig.~\ref{fig:sphist} panel d).
The actual filling factor may be larger than these values because of the effects of stray light
	\citep{2007PASJ...59S.837O}.
An upper bound can be computed by assuming the extreme case of $f=1$.
This yields a total vertical magnetic flux of $9.9\times10^{21}~\mx$, and a total horizontal magnetic flux of $2.0\times10^{22}~\mx$.
The difference is a factor of about $0.2$, which roughly corresponds to the average filling factor.

The total vertical magnetic flux for the whole area with latitude above $70^{\circ}$ is estimated to be between $5.6\times10^{21}$ and $2.5\times10^{22}~\mx$, assuming that the unobserved polar region has the same magnetic flux as the observed region.
Since the surface area with latitude above $70^{\circ}$ is $1.8\times10^{21}~\sqcm$, the average flux is estimated to be between $3.1$ and $13.9~\gauss$.
The total magnetic energy is proportional to $B^{2}f\,S = B\,\Phi$.
Thus, the surface poloidal magnetic energy is approximately two orders of magnitude larger than the case for the uniform magnetic field, if we take $B\sim1~\kgauss$, corresponding to the peak of the energy PDF in Fig.~\ref{fig:sphist}.
Though these are the most accurate flux estimation so far made for the polar regions, these number should be regarded as minimum values due to the threshold in the selection of pixels for accurate inversion.

From the Hinode observations, the total flux of vertical magnetic field at the polar region is estimated to be at least $5.6\times10^{21}~\mx$ and at most $2.5\times10^{22}~\mx$ at the solar minimum.
Various measurements indicate that the total magnetic flux of a single active region is about $10^{22}~\mx$
	\citep{2007SoPh..244...45L, 2007ApJ...671.1022J, mt08}.
Thus, the measured total polar flux barely corresponds to that of single active region.
The total toroidal flux would increase with time during the winding-up process by differential rotation, and the concept of the $\mathrm{\Omega}$-mechanism would be viable with these observational constraints.

\section{Bright points and magnetic elements}
\label{sec:bpme}
The magnetic field is found to be highly inhomogeneous in the lower solar atmosphere.
While field is likely ubiquitously present in the photosphere (cf.~Sects.~\ref{sec:thmf} and~\ref{sec:urmf}), it is concentrated at the edges of convective cells in small-scale regions of high field strength.
The convective flows expunge the field from cell interiors and concentrate the field in the intergranular downdrafts and at the borders of supergranular cells.
Field is concentrated in small ``magnetic elements'' that reach field strengths well beyond the equipartition field strength of about $500~\gauss$.
It should be noted that these elements are not discrete structures as their name suggests.
Rather, they are concentrations of strong field, with intricate structure that is expected to extend beyond what is already visible in observations at the highest spatial resolution.
In addition, they frequently split into multiple apparently disjoint concentrations, or merge with other concentrations during their lifetime.

The plasma $\beta$ in the photosphere outside kilogauss-strength magnetic elements, i.e., the ratio of the gas pressure to the magnetic pressure, is much larger than one.
In addition, because photospheric plasma has a high conductivity, the field is ``frozen in'' the matter.
As a result, the dynamics and evolution of magnetic fine structure in the photosphere are largely dominated by gas motions such as convection and large-scale flows associated with supergranulation.
Concentrations of strong magnetic field provides an excellent conduit for conveying kinetic energy from the turbulent photosphere to higher layers of the solar atmosphere.
If we are to understand the heating of the chromosphere and corona, as well as energetic events such as flares, it is important that we study the foot points of the field in the outer atmosphere.

\subsection{Observations of small-scale field concentrations}

\begin{figure}[tbp]
\begin{center}
\includegraphics[width=\textwidth]{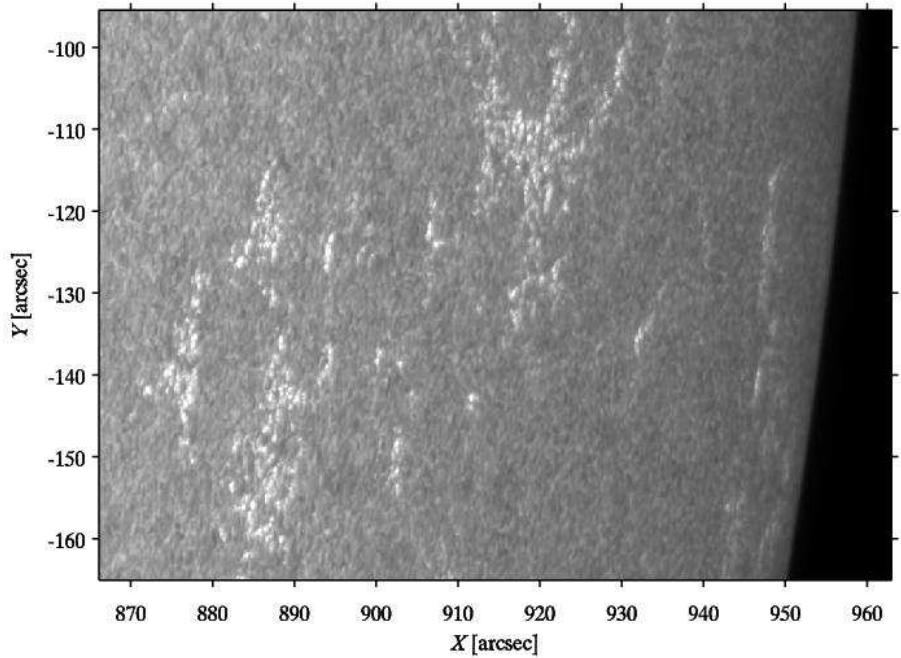}
\caption{A sample region in plage taken in the Fraunhofer G band near the west limb.
Faculae appear as small, bright features on the disk-center-side of granules.
This image was taken with Hinode/SOT on November~24, 2006, at 07:04:21~UT.}
\label{fig:faculaesample}
\end{center}
\end{figure}

There is a rich history of observations of concentrations of field in the solar photosphere.
Figure~\ref{fig:faculaesample} shows the most conspicuous small-scale magnetic features: faculae.
They show up as small bright features at the limb, usually in plages or decaying active regions.
For as long as the sun has been observed through telescopes, the existence of faculae has been known.
The counterparts of faculae closer to disk center are not as obvious in white light, but they do stand out in chromospheric diagnostics such as \CaIIH.

Small, concentrated magnetic elements in the network were observed as ``gaps'' in photospheric lines around $525~\nm$ by
	\cite{1967SoPh....1..171S},
and as ``magnetic knots'' in spectra of plage by
	\cite{1968SoPh....4..142B}.
It was clear from their observations that these structures were abundant in the vicinity of sunspots, but much more rare in quiet sun.
In wide-band \Halpha\ images, bright features were observed to be arranged in ``filigree'', a long-lived, large-scale photospheric network
	\citep{1973SoPh...33..281D}.
Observations by
	\cite{1974SoPh...38...43M}
had adequate resolution to resolve the photospheric network into strings of small ``bright points'' located in intergranular lanes, that change in shape and size on timescales comparable to the lifetime of granules.
The photospheric network had been associated with kilogauss field in the magnetic network earlier
	\citep{1973SoPh...32...41S},
suggesting that many of the observed structures were related
	\citep{1977SoPh...52..249M}.
Direct evidence that gaps, magnetic knots, faculae, filigree, and bright points were all manifestations of the same phenomenon was eventually provided by high-resolution observations, simultaneous in multiple wavelengths, of both disk and limb targets
	\citep{1981SoPh...69....9W}.
The bright points form a dense pattern in plage and active network, while outside of active regions, they clump in patches that partially outline supergranular cells.
The term ``network bright point'' was introduced to replace ``facular points'' and other terms in an effort to differentiate between bright points in regions of active and quiet sun
	\citep{1985SoPh...95...99S,
	1985SoPh..100..237M}.

\begin{figure}[tbp]
\begin{center}
\includegraphics[width=\textwidth]{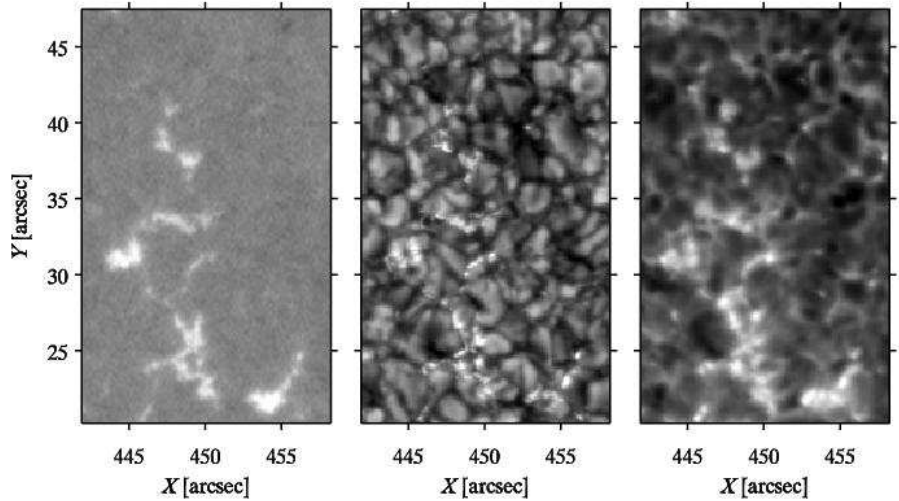}
\caption{Sample network region.
Left: \FeI\ $630.2~\nm$ line-of-sight magnetogram, scaled between $-1$ (black) and $1~\mathrm{k}\mxsqcm$ (white).
Middle: G-band intensity.
Right: \CaIIH\ intensity.
Bright points in the G band and in the \CaIIH\ line images correlate well with positions of concentrated field.
The network consists of strings of several adjacent bright points, located in the intergranular lanes.
Bright points in the \CaIIH\ image appear more extended and fuzzy than in the G-band image due to expansion of the flux tube with height.
These images were taken with Hinode/SOT on March~30, 2007, around 00:24:30~UT.
The coordinates indicate distance from sun center, so that $\mu\approx0.88$.}
\label{fig:proxysample}
\end{center}
\end{figure}

The importance of ``proxy-magnetometry'', the technique of determining the locations of magnetic field through a change in intensity, was recognized early on, and extensive studies of bright points in the photospheric network were quickly undertaken.
Imaging in wide-band \CaIIH\ and K or \Halpha\ were the diagnostics of choice, until
	\cite{1984SoPh...94...33M}
switched to using the Fraunhofer G band around $430.8~\nm$ in order to reduce the effects of chromatism in their telescope.
Imaging in the G band has since caught on and is now widely used as one of the principal diagnostics for proxy-magnetometry.
Figure~\ref{fig:proxysample} displays an example of proxy-magnetometry using imaging in the G band and in the \CaIIH\ line, together with a photospheric line-of-sight magnetogram.
The images show a small patch of network that consists of many bright points.

Imaging at consistently high resolution at high cadence over a reasonable duration is required to study the dynamics and evolution of magnetic elements, but it not easy to achieve.
Only during times of excellent seeing can an observer expect to make out these structures.
Thankfully, advances in digital imaging technology, the advent of adaptive optics, and the development of sophisticated algorithms for correction of the effects of atmospheric seeing in post-processing now allow telescopes to produce diffraction-limited data with some regularity.
In particular, many observers have successfully used observations with the former Swedish Vacuum Solar Telescope and the new 1-m Swedish Solar Telescope to study bright points, including size, shape, and appearance
	\citep{1995ApJ...454..531B,
	2004A&A...428..613B},
dynamics
	\citep{1996ApJ...463..365B,
	1998ApJ...509..435V,
	2005A&A...435..327R},
dispersal
	\citep{1998ApJ...506..439B},
and contrast
	\citep{2007ApJ...661.1272B}.
In addition, the space-borne observatory Hinode does not suffer from seeing and has produced a vast amount of data that is highly suitable for studies of network and internetwork field.

The relation of bright points to the underlying magnetic field has also received a fair share of attention.
Motivated by the relative ease with which data could be collected, observers often choose to study magnetic elements using proxy-magnetometry.
Comparison of diagnostics such as imaging in the G band with magnetograms has shown clearly that strong, kilogauss field is required to form a bright point, but it is not a sufficient condition
	\citep{2001ApJ...553..449B,
	2007A&A...472..911I}.
Many small concentrations of field that reach kilogauss strength do not have associated bright points, since the formation of the bright point depends strongly on the inclination of the field.
The contrast of a bright point decreases as the field is angled further away from the line of sight
	\citep{2007A&A...472..607B}.
The hope is, of course, that the observed bright points are a random sample of the magnetic elements, because field orientation is independent of the line of sight.
The results derived from these bright points are then expected to be valid for all magnetic elements, not just those that happen to have associated bright points, provided a statistically large number of bright points is sampled.
However, there are several reasons why one should be careful with these assumptions.
While proxy-magnetometry is comparatively simple, it is unable to continuously follow field concentrations if they are detected
	\citep{2005A&A...441.1183D},
and results based on these techniques are thus not just biased toward those concentrations that produce bright points, but also to the properties of those concentrations at the time that they are correctly angled to produce bright points.
Proxy-magnetometry misses a significant portion of flux that never becomes sufficiently concentrated to produces bright points, and is insensitive to field polarity.
It is important that results found through proxy-magnetometry be validated against measurements using a direct diagnostic of magnetic field.

\begin{figure}[tbp]
\begin{center}
\includegraphics[width=88mm]{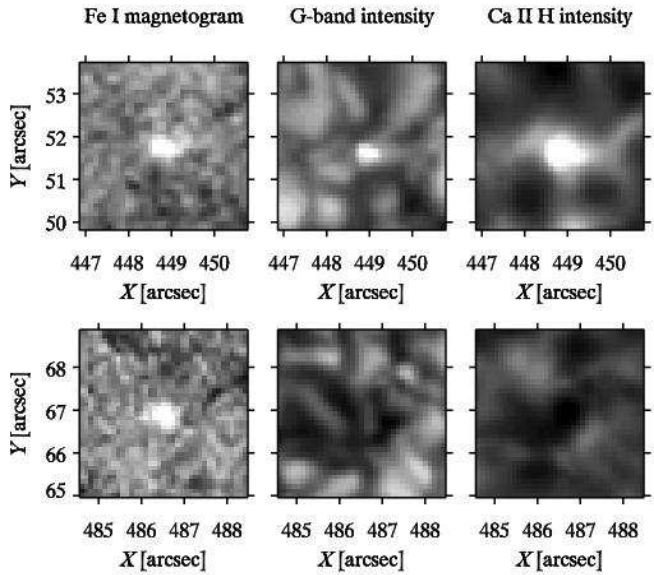}
\caption{Sample magnetic elements with (top row) and without associated bright points (bottom panel).
These magnetic elements appear similar in the line-of-sight magnetogram, yet are very different in G-band and \CaIIH\ intensity.
From the same sequence as Fig.~\ref{fig:proxysample}.}
\label{fig:globsamples}
\end{center}
\end{figure}

Figure~\ref{fig:globsamples} shows two illustrative examples of isolated magnetic elements in line-of-sight magnetograms and proxy-magnetometry diagnostics.
The elements appear similar in the magnetogram, yet only one has clear associated bright points in both G-band and \CaIIH\ filtergrams.
The other does not produce bright points, perhaps because the field is sufficiently angled away from the line of sight, or perhaps because the granulation around it is broken up and there is not enough proximity of hot granular walls to make a bright point.
These magnetic elements can exhibit different dynamics, as the one in the bottom row may not be buffeted by granulation as much as the one in the top row.
A study using a proxy-magnetometry diagnostic may therefore give different results than one based on true magnetometry.

\subsection{Field concentrations in internetwork areas}

Kilogauss fields that produce bright points can also be found in internetwork areas.
Though their existence was already noted in early studies of bright points
	\citep{1983SoPh...85..113M},
these fields have been largely ignored historically, likely because bright points are more isolated in the internetwork and thus much harder to identify.
There are fewer of them, and those that do exist are also more dynamic and have shorter lifetimes
	\citep{2003ApJ...587..458N,
	2005A&A...441.1183D}.
These factors make detection and analysis difficult.
Recently, however, vigorous investigation of magnetic field in internetwork areas has been undertaken.
Many of the observations used in these studies are now available thanks to the development of new instruments, adaptive optics, and post-processing techniques.
Analysis of the distribution of field strength and filling factor in the photosphere has attracted particular attention
	\citep{2003ApJ...582L..55D,
	2004ApJ...613..600L,
	2004Natur.430..326T,
	2006ApJ...636..496D}.
While there is some disagreement between results, weak field appears to be ubiquitously present in the internetwork, structured on small spatial scales (see Sect.~\ref{sec:mfsoqsf}).
In addition, horizontal field appears to pervade the photosphere (see Sects.~\ref{sec:hfiti} and~\ref{sec:thmf}).
New observations at unprecedented resolution with the space-borne observatory Hinode have shown that it is transient in nature, structured on small scales, and occurs preferentially over the edges of granules rather than in the intergranular lanes as is the case for more vertical field.
	\citep{2008ApJ...672.1237L},
These observations also allow us to study emergence of field on small spatial and temporal scales
	\citep{2007ApJ...666L.137C,
	2008A&A...481L..25I}.
Field is brought up inside a convective cell, then quickly expelled from the interior and swept into the lanes, where it may merge with pre-existing field.
The entire process takes only a few minutes.

Several studies have focused on strong field that has associated bright points.
The upshot is that internetwork field may become sufficiently concentrated to produce bright points, similar to network bright points, but more dynamic and with shorter lifetimes
	\citep{2004ApJ...609L..91S}.
The associated field has a longer lifetime than the bright point.
The field exists before the bright point is formed and remains after it disappears, and may produce a bright point again at some later time
	\citep{2005A&A...441.1183D}.
The lifetime of a bright point does not have a bearing on the lifetime of the underlying flux.
Rather, it is a measure of the dynamics of the associated flux, i.e., how long the flux remains sufficiently aligned with the line of sight to produce a bright point, and also of the performance of the detection algorithm.

\begin{figure}[tbp]
\begin{center}
\includegraphics[width=88mm]{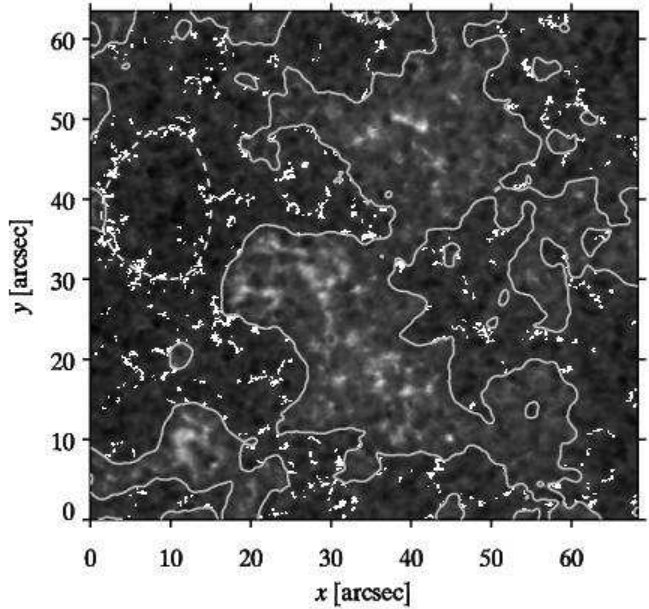}
\caption{\CaIIH\ intensity averaged in time over a 1-hour sequence.
The network is outlined by solid white contours.
Locations of internetwork bright points are overlaid in white.
The bright points appear to group in patches that outline edges of cell-like structures, such as around $(x,y)=(10\arcsec,40\arcsec)$ (indicated by a dashed line).
From
	\cite{2005A&A...441.1183D}.}
\label{fig:pmask}
\end{center}
\end{figure}

Strong field concentrations in internetwork areas also appear to outline cells on mesogranular scales.
Figure~\ref{fig:pmask} shows the locations of internetwork bright points detected in a 1-hour time sequence of \CaIIH\ images.
The interiors of these cells are largely devoid of field.
Similar patterns have been found in active network
	\citep{1998ApJ...495..973B}.
One would expect such a pattern to be set by granular motions.
Perhaps magnetic elements form these patterns as a result of flows associated with ``trees of fragmenting granules''
	(\citealp{2004A&A...419..757R},
	previously called ``active granules'' by \citealp{2001SoPh..203..211M})
which were previously linked to mesogranules
	\citep{2003A&A...409..299R}.
Flux is expunged by the sideways expansion of granular cells, and is collected in the downflows in intergranular lanes.
In a ``tree of fragmenting granules'', these flows would be expected to drive flux not only to the edges of individual granules, but also to the edges of the tree, resulting in a mesogranular pattern in the positions of magnetic field in internetwork areas.

\subsection{Magnetic element dynamics}

\begin{figure}[tbp]
\begin{center}
\includegraphics[width=\textwidth]{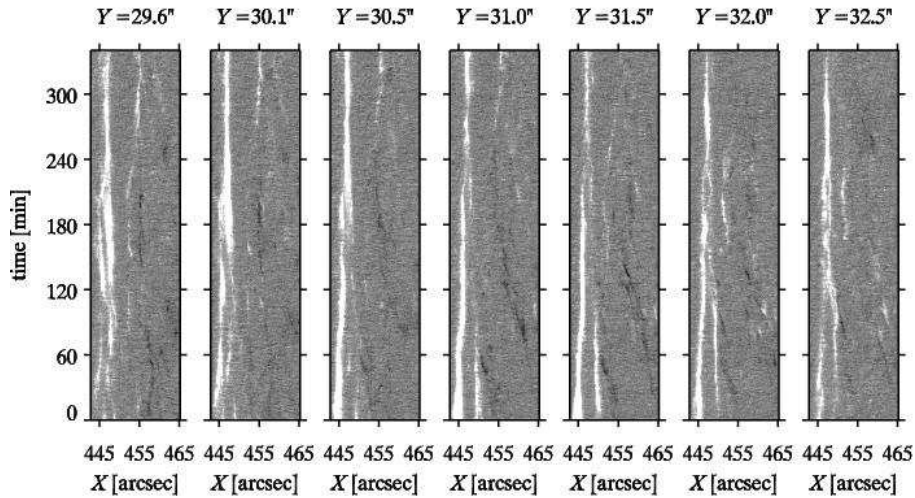}
\caption{Slices through a sequence of \FeI\ $630.2~\nm$ line-of-sight magnetograms, scaled between $-400$ (black) and $400~\mxsqcm$ (white).
The field of view includes some network at the left.
Magnetic elements exhibit dynamic behavior, migrating distances of several arcseconds over a few hours, while experiencing many interactions with other elements during that time.
They frequently seem to appear without a clearly associated opposite polarity.
Examination of adjacent slices indicates that such elements are not elements that emerged as a bipole previously and are now migrating into the current 2D slice.
This indicates that the flux emerged at some earlier time, and was only detected when it became concentrated enough, or that the associated flux of opposite polarity is spread out below the detection limit of the instrument.
From the same sequence as Fig.~\ref{fig:proxysample}.}
\label{fig:slice}
\end{center}
\end{figure}

Magnetic elements appear to obey largely Gaussian distribution of horizontal velocity, as would be expected from random buffeting by granulation.
Their rms velocity is about $0.5~\kms$ in network and about $1.5~\kms$ in internetwork.
Granular motions are suppressed in the network, resulting in less dynamic behavior.
Magnetic elements in internetwork areas sometimes migrate large distances over periods of a few hours (cf.~Fig~\ref{fig:slice}), while having frequent interactions with other long-lived elements and transient concentrations of field.
Magnetic elements do not typically have an identity over periods longer than a few minutes because of these interactions.

Motions of bright points and magnetic elements in internetwork areas show positive autocorrelation up to at least delay times of $10$~minutes, indicating that the elements retain some memory of their motions over at least that much time.
Magnetic elements in internetwork areas have motions preferentially in the direction of the nearest network concentration
	\citep{dewijn2008}.
The likely culprit is thus supergranular flow.

\subsection{Formation of bright points}

Modeling of strong magnetic concentrations began with analytic studies of magnetostatic ``flux tubes''
	\citep{1976SoPh...50..269S,
	1977PhDT.......237S}.
As computers became more powerful, numeric MHD models were created, increasing in complexity and realism over the years.
Early models were used to calculate properties of ``flux sheets'' in two dimensions
	\citep[e.g.,][]{1988A&A...202..275K}.
Modern three-dimensional numerical models of magneto-convection now simulate mesoscale-areas
	\citep[e.g.,][]{2006ApJ...642.1246S},
and are successful in reproducing magnetic elements and bright points in the solar photosphere
	\citep{2003ApJ...597L.173S,
	2005A&A...430..691S}.
However, two-dimensional models remain popular
	\citep{1998ApJ...495..468S},
because adding the third dimension is computationally expensive.

\begin{figure}[tbp]
\begin{center}
\includegraphics[height=42mm]{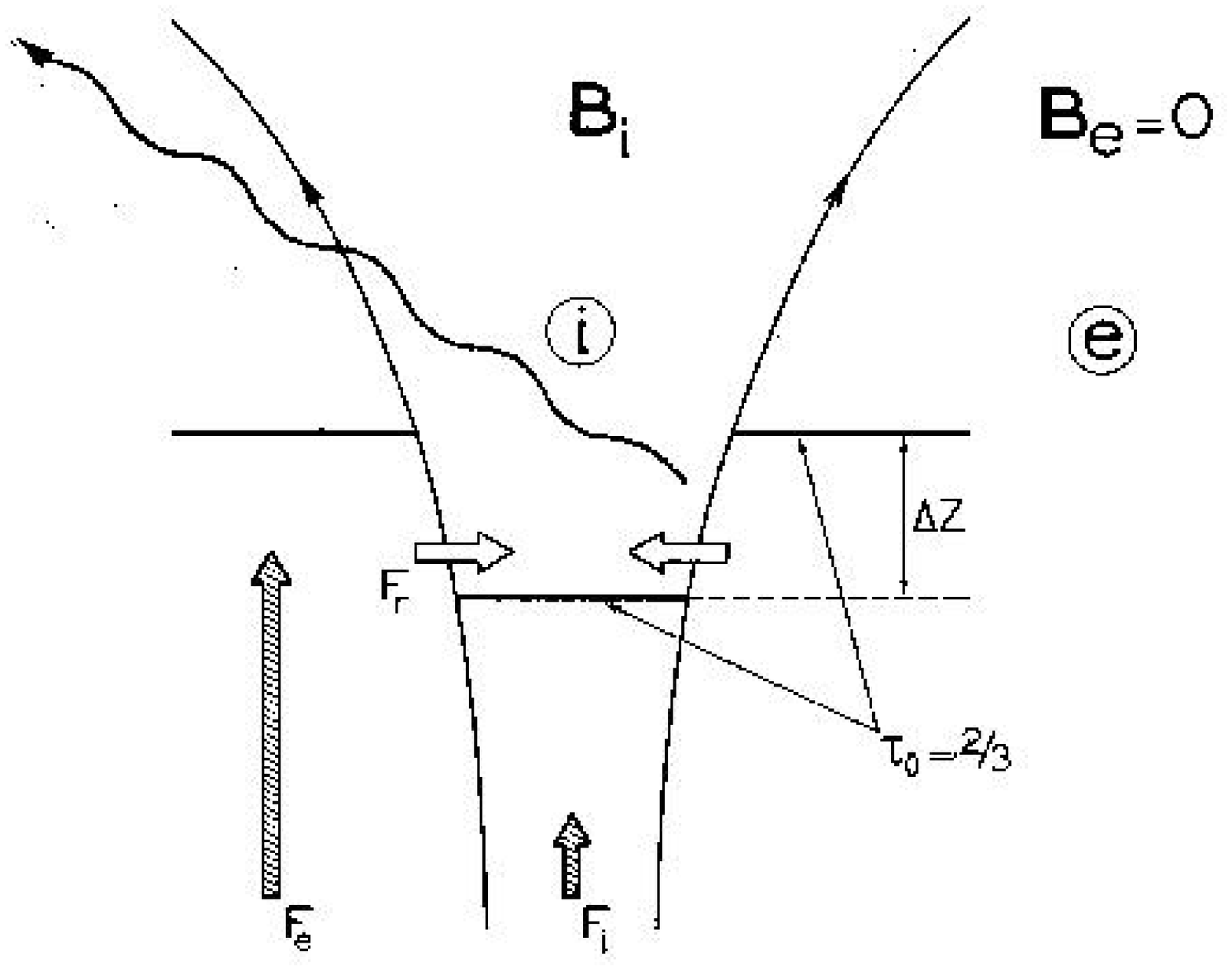}
\hfill
\includegraphics[height=42mm]{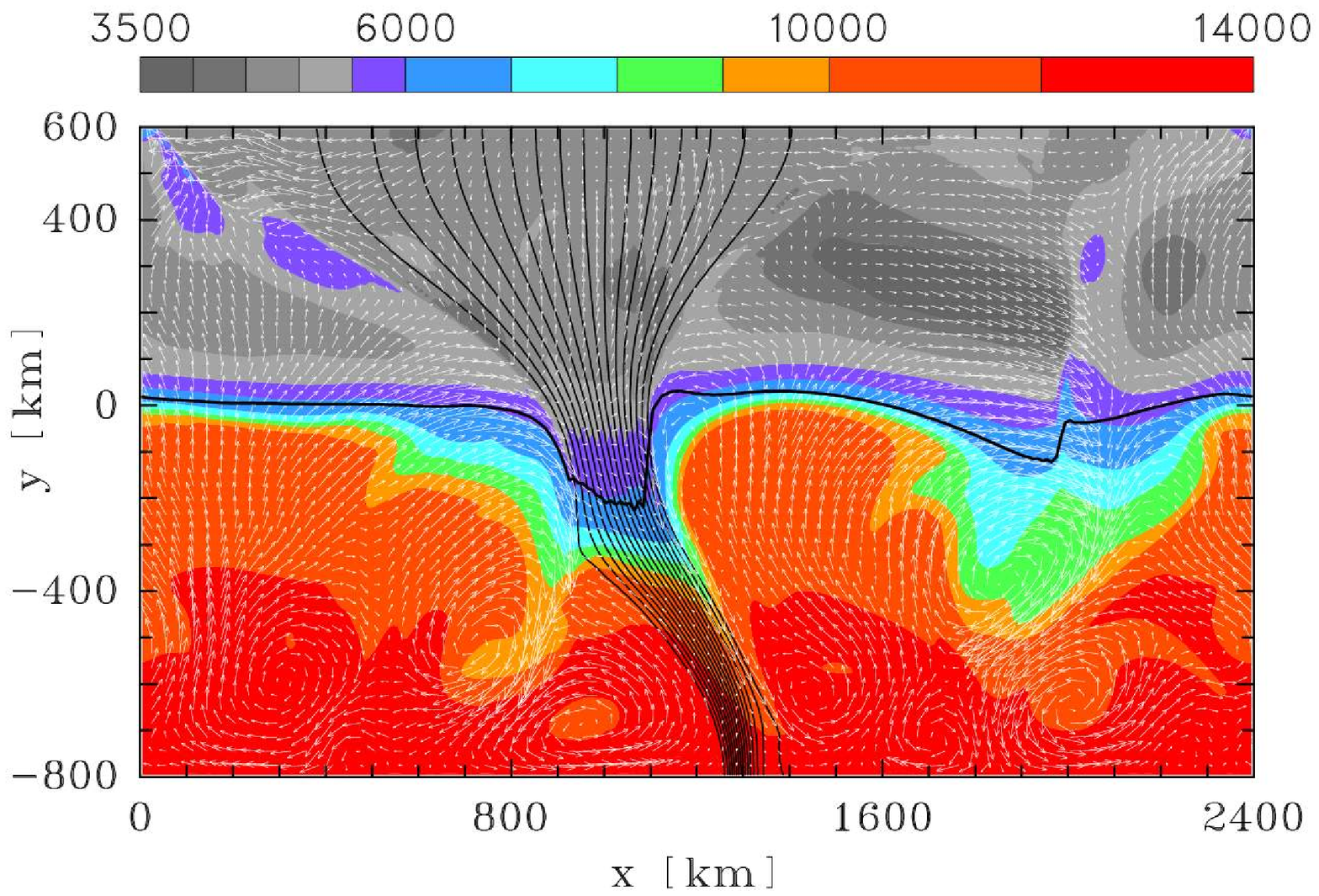}
\caption{Examples of modeling of magnetic elements.
Left: analytic magnetostatic flux tube model (from
	\citealp{cz2000}).
Right: sophisticated 2D numerical MHD model of a flux sheet (from \protect\url{http://www.kis.uni-freiburg.de/~steiner/}).}
\label{fig:fluxtubemodels}
\end{center}
\end{figure}

These models have shown that while the cause of brightness enhancement of magnetic elements in the photosphere differs subtly between various proxy-magnetometry diagnostics, it is in all cases rooted in the partial evacuation of the flux tube as a result of magnetic pressure.
The reduced density places optical depth unity inside the flux tube at a geometrically deeper layer compared to outside, thus allowing radiation to escape from deeper, hotter layers (cf.~the left panel of Fig~\ref{fig:fluxtubemodels}).
The internals of the flux tube are cooler at equal geometric height due to radiation losses, but are typically hotter at equal optical depth.
This process only produces enhanced brightness in small-scale structures.
Larger concentrations of strong field that inhibit convection, e.g., pores and sunspots, become dark because there is insufficient radial influx of radiation to make up for the increased losses as a result of reduced opacity.
Well-known diagnostics for proxy-magnetometry are used because they show more brightness enhancement than the continuum.
As an example, molecular lines such as those in the Fraunhofer G band and the CN band are weakened in small concentrations of field because of the dissociation of molecules at lower densities and higher temperatures
	\citep{2001IAUS..203..287K,
	2001A&A...372L..13S,
	2001ApJ...555..978S,
	2004A&A...427..335S,
	2006ApJ...639..525U}.
Opacity inside the flux tube is thus additionally reduced in these lines, and a higher emergent intensity integrated over the passband results.

One recent highlight was the confirmation of the suggestion made by
	\cite{1981SoPh...70..207S}
that facular brightness enhancement is the result of radiation escape from hot granular walls.
	\cite{2004ApJ...607L..59K} and
	\cite{2004ApJ...610L.137C}
used sophisticated 3D MHD simulations to model flux tubes, then ``observed'' them as if close to the limb using intricate codes to calculate radiative transfer.
Partial evacuation of the flux tube allows the observer to look deeper into the hot granular wall than would be possible if the flux tube were absent.
The dark lane often observed on the disk-ward side of faculae is formed in the cool layers above the granules and inside the flux tube.

\subsection{Formation of magnetic elements}

The prevailing theory on the formation of magnetic elements incorporates a convective instability known as ``convective collapse''
	\citep{1978ApJ...221..368P}.
Field is brought up in granules and swept into the intergranular downdrafts.
Flux can accumulate until the magnetic energy density is roughly equal to the kinetic energy density of granular flows.
This yields an equipartition field strength of about $500~\gauss$, insufficient to produce a bright point.
The gas in these regions cools because convective energy transport is suppressed by the field.
The cool, dense gas enhances the intergranular downflow, so that the region is effectively evacuated by gravity and consequently compressed until the internal magnetic pressure is sufficiently increased so that the region is again in horizontal pressure balance with the outside.
Theoretical calculations indicate that magnetic elements with field strengths of $1$--$2~\kgauss$ result from this process
	\citep{1979SoPh...61..363S,1998A&A...337..928G}.

The formation of kilogauss field concentrations from weak turbulent flux can be studied from models.
Typically, a hydrodynamic model is run until it reaches a more-or-less relaxed state.
A constant vertical field is then added, and the simulation is allowed to evolve further.
While this obviously does not resemble what happens on the sun, the process of convective collapse does still occur and indeed has been observed in simulations of magneotconvection
	\citep{2005A&A...429..335V}.

It is harder to observe convective collapse on the sun.
Accurate (spectro)polarimetric observations with high resolution and reasonable cadence are required over some period of time.
Such observations are sensitive to seeing conditions, due to, e.g., long exposure times required for polarimetry.
The seeing-free Hinode observatory is an obvious candidate to provide suitable observations.
Indeed, formation of a kilogauss field concentration by convective collapse was recently observed using the Hinode spectropolarimeter and found to be in qualitative agreement with results from numerical simulations
	\citep{2008ApJ...677L.145N}.
These observations strongly support the model of convective collapse for the formation of kilogauss field concentrations.

\section{Unresolved magnetic fields}
\label{sec:urmf}
\subsection{Range of the unresolved scales}\label{sec:range}

Magnetoconvection in the rotating sun is the engine of the solar dynamo.
It is therefore central to our understanding of the origin of solar and stellar activity.
A main problem in modeling the solar dynamo is that magnetoconvection has such a tremendous dynamic range, about 8 orders of magnitude or more, as we will see below, while numerical simulations can handle only about 3 orders of magnitude.
While numerical simulations provide valuable insights, the theory needs to be guided by observations.

The magnetic-field observations refer to the surface layers (photosphere), while the properties of magnetoconvection vary with depth in the convection zone.
Still, the photosphere can serve as our magnetoconvective laboratory, where we can explore the underlying physics.
However, a major part of the magnetoconvective spectrum extends over scales that are too small to be resolved even with next-generation telescopes in any foreseeable future.

One may therefore question to what extent knowledge about the behavior of these small unresolved scales is really needed for understanding solar and stellar dynamos and magnetic activity.
The solar dynamo gives the impression of being governed by large-scale properties like Hale's polarity law, Joy's law, the emergence and dispersion of active-region magnetic flux, shearing by differential rotation, and meridional circulation, all of which take place in the spatially resolved domain.
In Sect.~\ref{sec:role} we will address the connection between the scales and the role of the smallest diffusion scales for the operation of the solar dynamo.

The upper end of the scale spectrum is naturally bounded by the size of the sun ($\approx10^6~\km$).
The lower end of the magnetic spectrum is reached when the turbulent motions are unable to tangle the field lines to produce magnetic structuring.
This decoupling between the plasma motions and the magnetic field happens when the frozen-in condition ceases to be valid, i.e., when the time scale of magnetic diffusion (field-line slippage through the plasma) becomes shorter than the time scale of convective transport.

The ratio between these two time scales is represented by the magnetic Reynolds number
\begin{equation}
R_m =\mu_0\,\sigma\,\ell_c\,v_c
\end{equation}
in SI units.
Here $\sigma$ is the electrical conductivity, $\ell_c$ the characteristic length scale, and $v_c$ the characteristic velocities.
$\mu_0=4\pi\times10^{-7}$.
When $R_m\gg1$ the field lines are effectively frozen in and carried around by the convective motions.
When $R_m\ll1$ the field is decoupled from the turbulent motions and diffuses through the plasma.

The magnetic structuring by magnetoconvection therefore ends at scales $\ell_\mathrm{diff}$ where $R_m\approx1$.
To calculate these scales we need to know how the characteristic turbulent velocity $v_c$ scales with $\ell_c$.
Such a scaling law is given in the Kolmogorov theory of turbulence.
In the relevant inertial range it is
\begin{equation}
v_c=k\,\ell_c^{1/3},
\end{equation}
where $k$ is a constant.
An estimate of $k\approx25$ can be obtained from the observed properties of solar granulation (\AA{}ke Nordlund, private communication).

Combining these two equations and setting $R_m=1$, we obtain the diffusion scale
\begin{equation}
\ell_\mathrm{diff}=1/(\mu_0\,\sigma\,k)^{3/4}.
\end{equation}

To evaluate this we need the expression for the Spitzer conductivity in SI units,
\begin{equation}
\sigma=10^{-3}\,T^{3/2}.
\end{equation}
This gives us
\begin{equation}
\ell_{\rm diff}=5\times 10^5/T^{9/8}.
\end{equation}
For $T=10^4~\kelvin$ (a rounded value that is representative of the lowest part of the photosphere or upper boundary of the convection zone), $\ell_\mathrm{diff}\approx15~\m$.

If we limit ourselves to order-of-magnitude estimates, we may say that magnetic structuring in the surface layers ends at scales of about $10~\m$.
As the upper bound of the magnetic scale spectrum is about $10^6~\km$, it follows that the magnetoconvective scale spectrum spans about 8 orders of magnitude in the sun's observable surface layers.
As the diffusion limit decreases with increasing temperature, it follows that the dynamic range of magnetoconvection increases, possibly by nearly two orders of magnitude more (down to diffusion scales of cm) as we go down in the convection zone and the temperature increases towards a million degrees.

It might be objected that the rather simplistic Kolmogorov scaling law is not very applicable in the highly stratified surface layers.
However, the photospheric scale height of typically $150~\km$ is about 4 orders of magnitude larger than the diffusion scales that we have derived.
Most of these scales are so small that they do not ``feel'' the stratification and therefore may behave in a way that is similar to isotropic Kolmogorov turbulence.

\subsection{Role of the smallest scales for the global dynamo}\label{sec:role}

The magnetic fields that we see at the surface of the sun have been produced by dynamo processes in the solar interior.
Lifted by buoyancy forces, the dynamo-produced fields emerge as bipolar regions into the visible photospheric layers, but with an emergence rate that is a steep function of scale size.
The large-scale bipolar regions, which represent active regions (AR) with sunspots, bring up about $10^{20}~\mx$ per day (solar-cycle average), enough to account for the observed accumulation of flux and the large-scale background magnetic field over the course of the 11-year activity cycle.
Going down in scale to the so-called ephemeral active regions (ER), the flux emergence rate goes up by two orders of magnitude, to $10^{22}~\mx$ per day.
Going down to the still smaller internetwork fields (IN), the emergence rate increases to $10^{24}~\mx$ per day, another two orders of magnitude \citep{1987SoPh..110..101Z}.
While the characteristic scales of AR~:~ER~:~IN are in proportion $25:5:1$ ($75\arcsec:15\arcsec:3\arcsec$), the emergence rates are in proportion $1:100:10\,000$.

With these high emergence rates the time scale for the turn-over or replenishment of the magnetic field pattern is not the solar cycle time scale but something much shorter.
The first realization of a short turn-over time scale came two decades ago from a study of the differential rotation properties of the magnetic pattern \citep{1989A&A...210..403S}.
When determining the proper motion of magnetic elements through cross-correlation techniques \citep{1983ApJ...270..288S}, a steep differential rotation law is found, which closely agrees with the law derived from Doppler measurements.
When instead we form time series of the magnetic field sampled at the central meridian and perform an autocorrelation analysis to determine the period it takes for the pattern to recur after one solar rotation (or any integer number of rotation periods), then a rotation law is found that is almost rigid \citep{1989A&A...210..403S}.
This dramatic difference between the cross-correlation and autocorrelation analyses can be naturally explained if the pattern replenishment time is much shorter than a rotation period, so that the ``recurring'' pattern is not actually recurring but is a new pattern that has emerged during the course of the rotation period.

The nearly rigid differential rotation law then does not represent the surface (in contrast to the steep differential rotation law), but reflects the differential rotation properties of the source region in the deep convection zone, from which the new surface fields emanate.

This behavior cannot be easily explained in terms of flux-redistribution models without high-latitude sources of new magnetic flux, like the model of
\cite{1987ApJ...319..481S},
which are based on meridional circulation and a smooth surface diffusion process.
In such models a quasi-rigid differential rotation law for the phase velocity of the magnetic pattern results, regardless of the lag used in the correlation analysis.
The observed lag-dependence of the pattern phase velocity with a steep differential rotation law for small lags would not occur without the continual supply of new magnetic flux from the sun's interior at high latitudes.
To avoid this contradiction between the flux-redistribution models and the observations,
\cite{1994ApJ...430..399W}
replaced the smooth diffusion in their model with a discrete random walk process on a supergranular lattice, as a means of producing discrete flux clumps at high latitudes from the old, smooth, redistributed flux.
These clumps would then drift according to a steep differential rotation law.
It is, however, questionable whether supergranular random walk can continually produce flux clumps of sizes larger than one arcmin (the spatial resolution used in the correlation analysis of
\cite{1983ApJ...270..288S}
that gave the steep differential rotation law), much larger than the size of supergranules.
A more natural explanation is that the magnetic pattern is really being replenished from the sun's interior on a time scale well below the solar rotation time scale.

Support for such a short pattern replenishment time has come from SOHO MDI magnetograms, revealing a ``magnetic carpet'' with a pattern turn-over time of $1.5$--$3$ days \citep{1997ApJ...487..424S,1998ASPC..154..345T}.

The problem with the high emergence rates is that they have to be matched by the flux \emph{removal} rates for a statistically stationary situation, otherwise the photosphere would quickly get choked with magnetic flux that is all the time injected from below.
It is however difficult to identify the process by which flux is removed.
This problem is generally avoided in dynamo models by letting opposite polarities mathematically cancel out when they are co-spatial.
However, such mathematical cancellation is non-physical, magnetic flux can only be destroyed by a reconnection process involving concentrated electric currents and Joule heating, and this can only occur fast enough if it takes place on the diffusion length scales (of order $10~\m$ in the photosphere).
This implies an extreme and highly efficient shredding of the flux elements down to these scales, something that takes place almost entirely in the spatially unresolved domain and which is therefore not directly observed.

Flux removal may occur in basically three different ways: (1) In situ cancellation of opposite magnetic polarities (reconnection).
(2) Flux retraction (reprocessing in the convection zone).
(3) Flux expulsion (with a possible role of CMEs).
Unfortunately, the relative contributions of these three processes are completely unknown.
How impervious is the solar surface to the dynamo-produced magnetic flux?
How ``leaky'' is the solar dynamo?
These are fundamental questions that still have no answers.
Similar questions may be asked about the magnetic helicity.

Although reconnection is only mentioned explicitly in connection with the in situ cancellation, both flux retraction and flux expulsion could not happen without cutting off the field lines through reconnection.
Therefore, also for these processes, the basic physics takes place at the diffusion length scales, down to which the flux needs to be efficiently shredded.
Without this shredding, the global dynamo would not be able to operate.

\subsection{Scaling behavior of the magnetic field pattern}\label{sec:scaling}

The resolved scales now cover a dynamic range of almost four orders of magnitude (from the global scales of $10^6~\km$ down to the neighborhood of $100~\km$), approximately half of the range of the magnetoconvective scale spectrum.
Already back in the 1960s, in the early days of solar magnetography, when the dynamic scale range covered by the observations was only about two orders of magnitude, it was clear that the sun's magnetic field is very fragmented or intermittent, but the degree of intermittency or the nature of the structuring was not known.
To get an insight into the hidden nature of the field it was necessary to develop indirect diagnostic techniques to overcome the resolution limit and derive intrinsic field properties that were not dependent on the quality of the telescopes used.

A similar situation is encountered in stellar physics, where we derive the physical properties of the stellar atmospheres although the stars remain unresolved point objects.
While crucial information on key field parameters like magnetic field strengths and filling factors can be obtained this way, we have no information on the unresolved field \emph{morphology}, and the results depend on the interpretative models used.
An exception is Zeeman-Doppler imaging of rapid rotators.
By necessity these models have to be idealized to limit the number of free parameters, and they need to be tailored to the type of diagnostics that we use.
Thus the Zeeman and Hanle effects are sensitive to very different parameter domains of the field, as we will see in Sect.~\ref{sec:diagnostics}.

The situation has improved dramatically during the last decades.
Advances in spatial resolution have significantly extended the dynamic range of the resolved scales, allowing us to get a glimpse of how the magnetic pattern scales as we zoom in on ever smaller scales.
Numerical simulations have given us insights into the nature and scaling behavior of magnetoconvection when we go beyond the resolution limit into the unresolved domain.
This allows us to get a better understanding of the nature of the field pattern and gives us guidance in the choice of the most realistic interpretative models to use to diagnose the spatially unresolved domain.

Until a few years ago the ``standard model'' of photospheric magnetic fields was that the basic building blocks are strong-field (mostly kilogauss) highly intermittent flux tubes occupying a small fraction of the photospheric volume, and that the space between these flux tubes is filled with much weaker and highly tangled (or ``turbulent'') fields.
We now realize that this ``two-component picture'' is mainly a product of the idealizations used when interpreting Zeeman and Hanle signatures of the spatially unresolved domain.
Instead the field appears to behave like a fractal.

\begin{figure}[tbp]
\begin{center}
  \includegraphics[width=\textwidth]{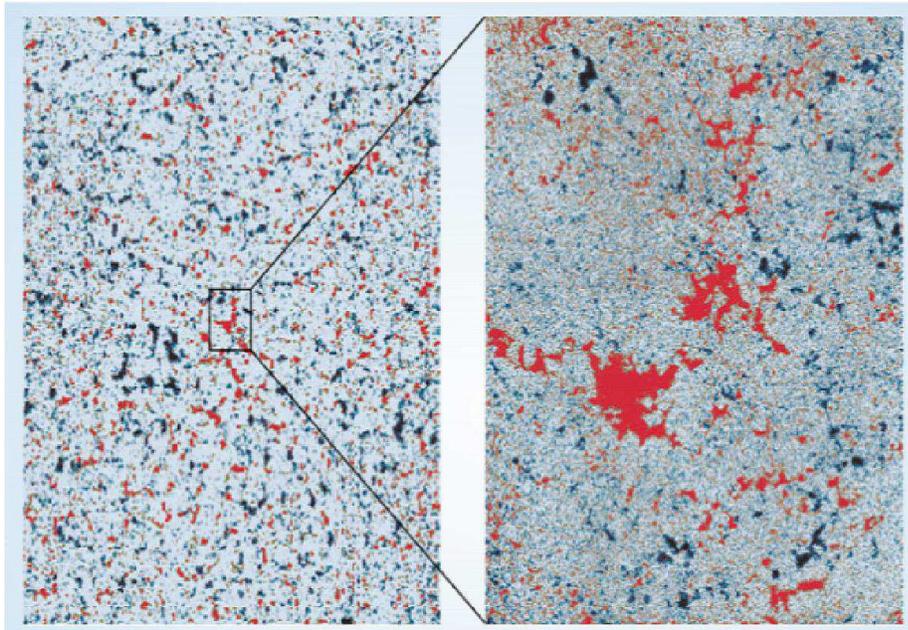}
\caption{Illustration of the fractal-like nature of the magnetic-field pattern on the quiet sun.
The left map, extracted from the central part of a Kitt Peak magnetogram of 9 February 1996, covers $15\%$ of the solar disk, while the right map, obtained on the same day at disk center with the Swedish La Palma telescope (courtesy G\"oran Scharmer) covers an area that is 100~times smaller \citep{2002ESASP.505..101S,2004Natur.430..304S}.}
\label{fig:fractal}
\end{center}
\end{figure}

Figure~\ref{fig:fractal} illustrates this fractal appearance of quiet-sun magnetic fields.
If a magnetogram is presented without tick marks that indicate the spatial scale, it is very hard to guess what the scale is.
The pattern seems to have a high degree of scale invariance, it looks statistically the same as we zoom in on ever smaller scales.
Further we have a coexistence of strong and weak fields over a large dynamic field strength range.
The probability distribution function for the field strengths appears to be nearly scale invariant and can be described in terms of a Voigt function with a narrow Gaussian core and ``damping wings'' that extend out to the kilogauss values (\citealp{2002ESASP.505..101S,2003ASPC..286..169S}, but see also \citealp{2006ApJ...636..496D}).
Such scale invariant properties are typical of a fractal.
A fractal dimension of $1.4$ has been found for both the observed magnetic field pattern and the pattern that results from numerical simulations of magnetoconvection at scales that are smaller than the resolved ones \citep{2003A&A...409.1127J}.

\subsection{Field diagnostics beyond the spatial resolution limit}\label{sec:diagnostics}

Like in any other area of astrophysics where we are dealing with spatially unresolved objects, we have to extract information about the physical conditions that is encoded as various types of signatures in the spectrum.
To enable this extraction we make use of models that must have a smaller number of free parameters than the number of independent observables that we can use to constrain the model.
A fundamental issue is the uniqueness and numerical stability of any such inversions.

\subsubsection{Zeeman diagnostics and the line-ratio technique}\label{sec:lineratio}

Back in the 1960s it became clear that the measured field strengths on the quiet sun increased with the spatial resolution of the instrument, which led to the question what the strength would be if we had infinite resolution \citep{1966ArA.....4..173S}.
To answer this question the line-ratio technique was devised, which led to the conclusion that more than $90\%$ of the net magnetic flux in the photosphere, as seen with modest spatial resolution (larger than a few arcseconds), comes from highly bundled fields with a strength of $1$--$2~\kgauss$ and a small volume filling factor (typically $1\%$) \citep{1972SoPh...22..402H,1972SoPh...27..330F,1973SoPh...32...41S}.
Due to the tiny filling factor the average net field strength is only of order $10~\gauss$ or less, although most of the field lines come from kilogauss flux patches in the photosphere.

This result led to the concept of discrete magnetic flux tubes as the theoretical counterpart of the unresolved kilogauss flux fragments.
The mechanism of convective collapse \citep{1978ApJ...221..368P,1979SoPh...61..363S,1979SoPh...62...15S} gained wide acceptance as the process leading to the spontaneous formation of kilogauss flux tubes.
Empirical flux tube models at increasing levels of sophistication were built \citep{1993SSRv...63....1S}.
Observational support for the convective collapse mechanism could be found \citep{1996A&A...310L..33S}, while also showing the existence of a family of weaker flux tubes that had been theoretically predicted \citep{1986Natur.322..156V}.

\begin{figure}[tbp]
\begin{center}
  \includegraphics[width=\textwidth]{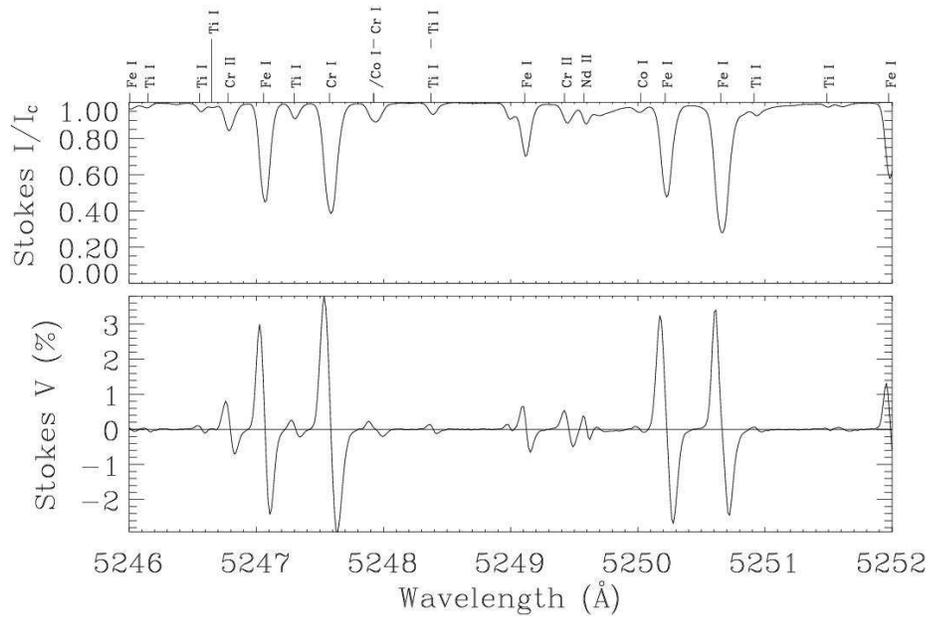}
\caption{Portion of a recording in a facula at disk center with the Fourier Transform Spectrometer at the McMath-Pierce facility (Kitt Peak) \citep{1984A&A...131..333S}.
The classical line-ratio technique is based on the comparison between the Stokes $V$ amplitudes in the two \FeI\ $524.7$ and $525.0~\nm$ lines \citep{1973SoPh...32...41S}.
Zeeman saturation due to strong, unresolved fields leaves a characteristic signature in the profile ratio, which allows the intrinsic field strength of the unresolved fields to be determined.}
\label{fig:zeeman}
\end{center}
\end{figure}

The classical line-ratio technique that allows a robust determination of the intrinsic field strength is based on the simultaneous observation of the circular polarization in the \FeI\ $524.7$ and $525.0~\nm$ line pair \citep{1973SoPh...32...41S}.
The observed circular polarization due to the longitudinal Zeeman effect, illustrated in the FTS spectrum of Fig.~\ref{fig:zeeman}, depends on many combined factors, like the line depth and detailed line shape (which in turn depend on the temperature-density stratification of the atmosphere and the details of line formation), the Land\'e factor, and the line-of-sight component of the magnetic field.
In traditional magnetography one calibrates away the line-profile factors by recording the magnetograph response to artificial line shifts of the spatially averaged spectral line.
This calibration procedure then gives us the average line-of-sight field strengths (averaged over the spatial resolution element of the instrument), under the assumption that the relation between circular polarization and field strength is a linear one (being the relation that is valid in the weak-field limit), and assuming that the line depth and line shape are the same in the magnetic elements as for the spatially averaged sun.
However, both these assumptions are generally wrong.
As the field strength increases, the relation between circular polarization and field strength becomes increasingly non-linear.
Further, the temperature-density stratification of the atmosphere (and thus the line profile) is significantly different in the unresolved magnetic elements than outside them.
The line-ratio technique allows us to isolate the magnetic-field effects from the line formation and temperature-density effects, to obtain a signature that can only occur if the field is intrinsically strong (meaning that the polarization dependence on field strength lies in the non-linear regime).
From the measured degree of non-linearity (Zeeman saturation) the value of the field strength can be extracted.

The robustness of the method depends on the choice of line pair.
No line pair has been found that better optimizes this robustness than the \FeI\ $524.7$ and $525.0~\nm$ one.
These two lines both belong to multiplet no.~1 of iron, have almost identical excitation potential, oscillator strength, and line depth, and therefore respond in the same way to the temperature-density stratification of the atmosphere, with the same line formation properties.
The only significant difference between them is their effective Land\'e factors: $2.0$ for the $524.7~\nm$ line, $3.0$ for the $525.0~\nm$ line.
If all fields were intrinsically weak, the circular-polarization Stokes $V$ profiles of the two lines would have the same shapes and only differ in terms of a global amplitude scaling factor in proportion to their Land\'e factors ($g_{524.7}:g_{525.0}=2:3$).
If we form the ratio $g_{524.7}\,V_{525.0}/(g_{525.0}\,V_{524.7})$, it would be unity if all fields were weak, regardless of the temperature-density stratification or line-formation properties of the solar atmosphere.
It differs from unity only because of the \emph{differential non-linearity}: the $525.0~\nm$ line with its larger Land\'e factor deviates more from linearity than the $524.7~\nm$ line.

This line-ratio technique was applied before the advent of Stokesmeters, using magnetograph exit slits in fixed positions of the line profiles \citep{1973SoPh...32...41S}.
This was sufficient for obtaining robust field-strength determinations.
With fully resolved Stokes $V$ line profiles with high S/N ratio (cf.~Fig.~\ref{fig:zeeman}) it became possible to test and verify the interpretation in great detail, since the Zeeman saturation does not only suppress the Stokes $V$ amplitudes but also broadens the Stokes $V$ profile in a way that gives the $g_{524.7}\,V_{525.0}/(g_{525.0}\,V_{524.7})$ ratio a very characteristic profile shape when plotted as a function of wavelength $\mathrm{\Delta\lambda}$.
Thus the self-consistency and validity of the interpretational model could be verified (for details, see \citealp{stenflo1994}).

This interpretational model contained two components: one magnetic component with field strength and filling factor as the free parameters, and one non-magnetic component.
The measured line ratio does not depend on filling factor, only on field strength.
The filling factor enters when explaining the $V$ amplitudes of each line, since the amplitudes scale with both filling factor and field strength.
Since the line ratio was found to be practically identical in quiet network regions with little magnetic flux and in strong faculae with much flux, the conclusion was that the magnetic building blocks (flux tubes) have rather unique properties \citep{1972SoPh...27..330F}, almost always with field strengths of $1$--$2~\kgauss$.
Different regions on the sun (outside sunspots) then differ not so much in field strength, but rather in the number density or filling factor of the flux elements.
This implies that the magnetograms, which show a continuous range of apparent field strengths, basically are maps of the filling factor, not of field strength.

This view of solar magnetism has been confirmed with other combinations of spectral diagnostics, in particular with infrared lines (e.g., \citealp{1992A&A...263..323R}), which however have also revealed the existence of intrinsically weaker flux elements that are mixed in with the kilogauss ones.
Due to the larger Zeeman splitting in the infrared it was possible to extend the 2-component approach to a 3-component one (with two magnetic components), which revealed the existence of intrinsically weaker fields.
With advances in spatial resolution it became possible to actually resolve and see the flux tubes that had been predicted by the line-ratio method, as first done with speckle polarimetry \citep{1992Natur.359..307K}.

\subsubsection{Hanle diagnostics}\label{sec:hanlediag}

While all these results were self-consistent, the Zeeman-effect observations left us with a picture where about $99\%$ of the photospheric volume (outside the kilogauss flux elements) was field free, which is non-physical, since nothing in the highly electrically conducting and turbulent photospheric plasma can be field free.
The introduction of a ``non-magnetic'' component is exclusively for mathematical convenience.
The question is what the magnetic nature of this component is.
Since its contribution to the Zeeman-effect polarization signals is very small, it must either mean that the field is indeed extremely weak, or that the field is highly tangled with mixed polarities within the spatial resolution element, such that one has nearly perfect cancellation of the opposite signs of the spatially unresolved Stokes $V$ signals (in which case the field does not have to be weak).
We now know through applications of the \emph{Hanle effect} that the second case is much closer to the truth.

\begin{figure}[tbp]
\begin{center}
  \includegraphics[angle=-90.,width=\textwidth]{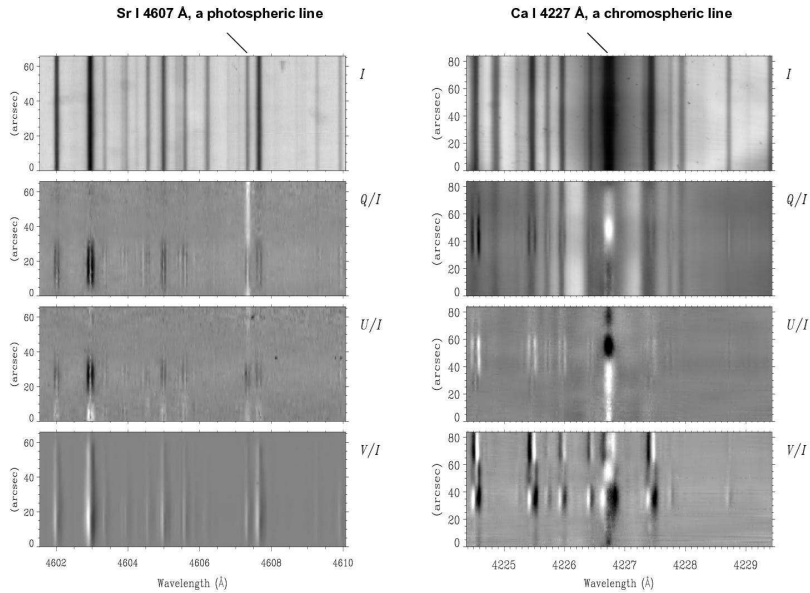}
\caption{Illustration of the different signatures of the Zeeman and Hanle effects in images of the Stokes vector (represented by the images of the intensity $I$ and the three fractional polarizations $Q/I$, $U/I$, and $V/I$).
The Hanle effect appears in the linear polarization (Stokes $Q/I$ and $U/I$) in the line cores of certain lines, like \SrI\ $460.7~\nm$ (left panels) and \CaI\ $422.7~\nm$ (right panels), while the Zeeman effect exhibits its usual polarization signatures in the surrounding lines.
The recordings were made with the Zurich Imaging Polarimeter (ZIMPOL, cf.~\citealp{1995OptEn..34.1870P,2004A&A...422..703G}) at the McMath Pierce facility (Kitt Peak).}
\label{fig:hanle}
\end{center}
\end{figure}

In contrast to the Zeeman-effect polarization, the Hanle effect is a coherency phenomenon that only occurs when coherent scattering contributes to the line formation.
Such scattering can produce linear polarization also in the absence of magnetic fields.
The term Hanle effect covers all the magnetic-field induced modifications of the scattering polarization.
Since it has different sensitivity and symmetry properties than the Zeeman-effect polarization, it both responds to much weaker fields and does not suffer from the cancellation effects that make the Zeeman effect ``blind'' to a tangled field.
This property was first exploited by \cite{1982SoPh...80..209S} to derive a \emph{lower} limit of $10~\gauss$ for the strength of the tangled field in the $99~\%$ of the volume between the kilogauss flux tubes.

Examples of Hanle-effect signatures and how they differ from the Zeeman effect are shown in Fig.~\ref{fig:hanle}.
The photospheric \SrI\ line in the left panels has been extensively used by various authors \citep{1993A&A...268..765F,1995A&A...298..289F,1998A&A...329..319S,2004Natur.430..326T} to improve the constraints on the properties of the turbulent field for which the Zeeman effect is blind.
The most sophisticated constraints based on the use of probability distribution functions (PDF) have been derived by \cite{2004Natur.430..326T}, indicating turbulent field strengths of order $100~\gauss$.
Such volume-filling fields contain so much magnetic energy that they may play a major role in the energy balance of the solar atmosphere.
The Hanle signatures of the strong \CaI\ $422.7~\nm$ line (right panels in the figure) can be used to diagnose the horizontal magnetic fields in the solar chromosphere.

Assume that we have chosen our Stokes coordinate system such that the non-magnetic scattering polarization is along the Stokes $Q$ direction.
This direction is parallel to the nearest solar limb when observing on the solar disk in a zone near the limb (which we most often do for such observations, since the scattering polarization amplitude increases as we get closer to the limb).
The main polarization signatures of the Hanle effect are depolarization (reduction of the Stokes $Q$ amplitude) and rotation of the plane of polarization (appearance of signals in Stokes $U$).
Since however the Hanle rotation angle can have both signs, a highly tangled field with equal contributions of plus and minus will lead to cancellations like for the Zeeman effect, so there will be no Stokes $U$ signatures from such fields.
In contrast, the depolarization effect in Stokes $Q$ has only one ``sign'' (reduction of the polarization amplitude), regardless of the field polarity, and is therefore immune to the above-mentioned cancellation effects.
This is the signature of the turbulent fields that we have to work with.

For Hanle diagnostics of the turbulent fields Hanle depolarization gives us one observable per spectral line.
For single-line observations, the interpretative model therefore cannot contain more than one free parameter.
Since the introduction of this diagnostic technique, the traditional model has been to assume a single-valued field with an isotropic angular distribution \citep{1982SoPh...80..209S}.
The free parameter is the single-valued field strength.
However, other more realistic model choices are beginning to be used, which are guided by the insights gained from numerical simulations of magnetoconvection and from analysis of magnetic-field distribution functions that have been determined from observations in the spatially resolved domain.
From the observed scaling behavior of the resolved fields and the behavior of the smaller-scale fields in numerical simulations one can make educated guesses for the analytical shapes of the field strength probability distribution functions (PDFs) that should be used to model the behavior in the spatially unresolved domain.
With a single Hanle observable (the depolarization for a single spectral line) we must then limit ourselves to characterize the model PDF with a single free parameter, for instance by keeping the relative shape invariant and using a stretching factor as the free parameter.
Such an approach has been applied with success for data with the \SrI\ $460.7~\nm$ line \citep{2004Natur.430..326T}.

The Hanle effect however does not at all limit us to use such simplistic, one-parameter models.
They are only used because the application of the Hanle effect to the diagnostics of spatially unresolved magnetoconvection is still in its infancy, and one needs to start with and fully understand the simplest approaches before proceeding to higher levels of sophistication.
Like with the line-ratio technique in the case of the Zeeman effect, with the Hanle effect one can also use a multi-line approach with simultaneous Hanle observations in several spectral lines with different sensitivities to the Hanle effect.
Such an application of the \emph{differential Hanle effect} \citep{1998A&A...329..319S} increases the number of independent observables, which allows us to increase the number of free parameters of the interpretative models and thus enhance the degree of realism.
While most lines differ not only in their Hanle sensitivities but also in their line formation properties, which adds considerable complication to the inversion problem, there exist certain pairings of molecular lines for which the line formation properties are identical, the only difference being the Hanle sensitivities \citep{2004A&A...417..775B}.
This allows the magnetic-field effects to be isolated from the other non-magnetic effects, similar to what is done with the $525.0$/$524.7$ Zeeman-effect line ratio.
This has the great advantage of making the inversion much more robust and the derived field strengths less model dependent.

\subsubsection{Unified Zeeman-Hanle diagnostics with distribution functions}\label{sec:unified}

The Zeeman and Hanle effects are highly complementary.
The longitudinal Zeeman-effect signals represent the net magnetic flux that often (but not always) has its main sources in the highly bundled strong fields, but they carry nearly zero information on the spatially unresolved volume-filling weaker, tangled fields between the intermittent stronger fields. Let us here recall that nearly four orders of magnitude in spatial scales lie unresolved below the current spatial resolution limit of magnetograms (as represented by Hinode, cf.~Sect.~\ref{sec:range}).
In contrast, the Hanle effect is almost blind to the flux-tube like fields, for three reasons: (1) The effect scales with the filling factor, which is very tiny for the flux-tube fields (of order $1\%$).
(2) The Hanle effect is insensitive to vertical fields, and the strong fields tend to be nearly vertical due to the strong buoyancy forces acting on them.
(3) The Hanle effect completely saturates for fields stronger than a few hundred gauss.
The complementary nature of the two effects has in the past led to the choice of two apparently contradictory interpretative models used for each effect: for the Zeeman effect the two-component  model (or extended variations thereof, with additional components) with the concept of a magnetic filling factor, for the Hanle effect a volume-filling field (filling factor of unity) with an isotropic distribution of field vectors.

This apparent dichotomy in the diagnostic methods arises because each of the Zeeman and Hanle effects provides an incomplete, filtered view of the underlying reality, which in a unified picture is fractal-like, and which may best be characterized in terms of probability distribution functions (PDFs).
When we ``put on our Zeeman goggles'', we project out properties of the strong-field tail of the PDF, which appears flux-tube like.
When, on the other hand, we ``put on our Hanle goggles'', we project out properties of the weak-field portion of the PDF.
However, the application of unified PDF models for both Zeeman and Hanle diagnostics is still in its infancy, and the initial results are only tentative, because the information we have that could guide our choice of distribution functions is still very incomplete.

The incompleteness mainly lies in the lack of information on the angular distribution function of the field, not so much in the PDF for the field strengths, for which we have reasonably good analytical functions to work with.
The angular distribution is expected to be closely coupled to the field-strength distribution.
From theoretical considerations we expect the stronger fields to have an angular distribution that is fairly peaked around the vertical direction, since they are more affected by the vertical buoyancy forces while resisting bending and tangling by the turbulent motions.
The weakest fields on the other hand are expected to have a much wider angular distribution, since the dominating effect is the turbulent tangling of the passive fields.
For intermediate field strengths we should have a gradual transition between the wide and the peaked angular distributions.
The Stokes profile signatures from such combinations of distribution functions for the spatially unresolved magnetic fields have recently been explored by radiative-transfer modeling \citep{sampoornaetal08}, but such calculations have not yet been applied to model fitting of observational data.

A unique opportunity to obtain lacking observational information on the angular distribution functions would be with the SOT data from the Hinode spacecraft.
A detailed exploration of the distribution functions of the quiet-sun vertical and horizontal magnetic fields with Hinode data \citep{2008ApJ...672.1237L} has given the surprising result that there seems to be five times more horizontal magnetic flux than vertical flux.
Furthermore, the patches of flux concentrations of vertical and horizontal fields are observed to be well separated, rather than co-spatial.
There is convincing indirect evidence that most of the horizontal flux patches are not spatially resolved even with Hinode but have a small filling factor, indicating intrinsic sizes of the underlying flux elements of at most $50~\km$ \citep{2008A&A...481L..25I}.
These intriguing results are not yet properly understood, so the angular distribution functions needed for our diagnostic models still remain elusive.

\section{Conclusion}
\label{sec:conclusion}
While we have attempted to give a comprehensive overview of small-scale magnetic field in the solar atmosphere, this review is by no means complete.
In particular, we have not touched upon chromospheric fields, which besides being structured on small scales, also display dynamic behaviour on short timescales.

New instruments that are able to measure photospheric magnetic field in high-resolution and at sufficient cadence to study dynamics, either directly through (spectro)polarimetry, or indirectly through proxy-magnetometry, are now availble to observers.
In particular, we have discussed several results from the Hinode mission.
These results, while dealing with small-scale field, have great repercussions for important questions surrounding magnetism in the sun, and in particular for the existence and workings of both the local and global solar dynamos.

With new instruments and sophisticated modeling enabled by advances in computing, we have greatly improved our understanding of magnetic activity on all scales in the sun.
Yet, we have also seen that the end is not yet in sight: field is likely structured on scales well beyond what can be observed or simulated today or in the forseeable future.
Our understanding of the processes that give rise to small-scale magnetic field will continue to improve as more observations are analyzed, models become more sophisticated and lifelike, and new instruments are developed, such as the Sunrise baloon-borne observatory
	\citep{Gandorfer2006}
or the Advanced Technology Solar Telescope
	\citep{Keil2000}.

\end{document}